\documentclass[bibyear]{aa}
\usepackage{caption}
\pdfoutput=1
\usepackage{graphicx}
\usepackage {subcaption}
\usepackage{float}
\usepackage{subfloat}
\makeatletter
\def\@bibitem#1{\item[]%
    \if@filesw\immediate\write\@auxout{\string \bibcite {#1}{\the\value{\@listc\
tr }}}\fi\ignorespaces}
\makeatletter

\begin{document}

\title{Spiral structure generated by major
planets in protoplanetary disks}
\subtitle{The role of periodic orbits near resonance}

\author {R.H.~Sanders}
\institute{Kapteyn Astronomical Institute,
P.O.~Box 800,  9700 AV Groningen, The Netherlands}
 \date{received: ; accepted: }

\titlerunning{Spiral structure in protoplanetary disks}

\abstract{

In this paper I describe numerical calculations of the motion of particles in a disk
about a solar-mass object perturbed by a planet on a circular orbit with
mass greater than 0.001 of the stellar mass.  A simple algorithm for simulating
bulk viscosity is added to the
ensemble of particles, and the response of the disk is followed for several
planet orbital periods.  A two-arm spiral structure forms near the inner  resonance
(2-1) and extends to the planetary orbit radius (corotation). In the same way for gaseous
disks on a galactic scale perturbed by a weak rotating bar-like distortion, this is shown to
be related to the appearance of two perpendicular families of periodic orbits
near the resonance combined with dissipation which inhibits the crossing
of streamlines. Spiral density enhancements result from the crowding of
streamlines due to the gradual shift between families.  The results,
such as the dependence of pitch-angle on radius and the asymmetry of the
spiral features, resemble those of sophisticated calculations that include
more physical effects. The morphology of structure  generated in this way
clearly resembles that observed in objects with well-defined two-arm spirals,
such as SAO 206462.
This illustrates that the process of spiral
formation via interaction with planets in such disks can be due to orbital motion
in a perturbed Keplerian field combined with kinematic viscosity.}

\keywords{
Planetary systems: protoplanetary disks -- planet-disk interactions -- methods: hydrodynamics
}
\maketitle
\section{Introduction}
A striking result of recent high resolution near-infrared and
submillimeter to millimeter observations of disks around young stars is the
discovery of various structures (Perez et al. 2016, Avenhaus et al. 2018):
rings, gaps, and spirals.  These disks, on a
scale of tens of astronomical units, are
the sites of nascent planetary systems, and the observed structures
may reflect, either as cause or effect, the formation of planets
within the disks.  Initially the detected structures
were observed primarily in the polarized near-infrared scattered light that
arises in the thin outer layers of the disks, and thus could
 essentially be a surface phenomenon (Dong et al. 2018). Detection
of structure in CO line emission (e.g., Tang et al. 2017) is also
relevant to the outer disk layers because of the large optical
depth in these spectral lines.  The more recent detection of
structure in the thermal dust emission at millimeter wavelengths
(Huang et al., 2018) more likely reflects the basic underlying structure of the disks
implying that the structure  is not merely surface phenomenon.  This is significant because
the mid-plane is the likely site of planet formation (a caveat is that the large
grains responsible for millimeter wave emission are more susceptible to aerodynamic
drag, and thus may not reflect the overall gas distribution; Andrews 2020).

\vskip 1\baselineskip
For decades it has been known that satellites can sculpt the
structure of
planetary ring systems (like that of Saturn) by   clearing gaps, forming
rings, and exciting spiral structure (Goldreich \& Tremaine 1979, 1980;
Shu 1984).
The new observations of protoplanetary disks (PPDs) may be evidence of a similar phenomenon on the scale
of forming planetary systems, as explored in the work of Goodman \&
Rafikov (2001) and Ogilvie \& Lubow (2002). Although most of the
structure observed is essentially axisymmetric (rings, gaps), a non-trivial
fraction, perhaps
one-fifth to one-fourth, is of spiral form,
in some cases even grand design two-arm spirals, as in
galaxy disks (Dong et al. 2018).
This suggests that the mechanisms for the generation of structure in
PPDs are similar to those in galaxy disks:
gravitational instability in the gaseous disk or the
effect of pre-existing non-axisymmetric
structure in the gravitational field
(or some combination of the two).
It is the second mechanism that I  consider here.

\vskip 1\baselineskip
In galaxies detailed structure is more evident in gas-rich
disk systems than in spheroidal systems or early-type disks
that are relatively
deficient in gas. An aspect of this difference
is due to star formation which, in effect, lights up the existing
structure in the disk. But the argument has been made
that the basic structure itself is due to the presence of
a dissipative medium that responds in a highly non-linear
way to weak perturbations in the axisymmetric
potential of the dominant stellar component:  a spiral or
a bar form in the underlying stellar distribution, and hence
in the gravitational field. For example, a weak bar distortion
in the potential can excite conspicuous rings or spirals
in the gaseous component (Sanders \& Huntley 1976, Schwarz 1981).
In this phenomenon periodic orbits
play a primary role (van Albada \& Sanders 1982)
in forming the basis for gas streamlines.
\vskip 1\baselineskip
The same must be true in PPDs, but in this case
the non-axisymmetric distortion is due to forming or formed
giant planets.  As in the case of galaxy disks it is near the
inner-Lindblad resonance (a test
particle orbits twice as the planet orbits once) where periodic orbits far from
the planet itself
have the largest deviations from circular motion and have their dominant effect
in initiating structure formation.  However, there are obvious differences with
general galactic disks:  in the case of a
planetary system the
perturbation is one-sided, corresponding to an m=1 distortion in
a Fourier description, but the response
of orbits  at the inner resonance is bisymmetric, m=2.  Thus,
the gas response near the resonance would be expected
to reflect this symmetry.
I argue here that the overlapping of two families of periodic orbits in a dissipative
medium is the basic driver of
two-arm spiral structure formation excited by planets in PPDs, although one would expect m=1 variations about this
bisymmetric structure near corotation where the planet is no
longer a weak perturbation.
\vskip 1\baselineskip
The formation of spirals in disks perturbed by massive planets
has been demonstrated in earlier numerical hydrodynamical calculations,
so this is not a new result (Dong et al. 2015;
Zhu et al. 2015). These
previous calculations are sophisticated, including effects such as
three-dimensional structures, multi-components, and radiative transfer,
whereas the simulations employed here are simple in comparison.
The simulations are two-dimensional, where
the zeroth-order axisymmetric force is the inverse square
law due to the central
star (on the order of one solar mass).  This symmetry is broken by a planet
of one-thousandth to several thousandths of the stellar mass (one to
several Jupiter masses) on a circular orbit within the disk at
a distance of several tens of
astronomical units.  The disk is represented by an ensemble
of several thousand particles, and the effects of dissipation are simulated
by giving each particle an interaction size
on the order of two astronomical units,  an interaction
which reduces the velocity differences between particles
over this distance.
\vskip 1\baselineskip
The goal is not primarily to model specific systems
(although one specific system with a well-defined two-arm structure
is considered), but
to reduce the problem to its essential elements.
The only physics entering the calculation is two-dimensional
motion in a perturbed Keplerian potential including a crude mechanism
for simulating viscosity.  Nonetheless, with this brutalized
approach the spiral structure revealed in more
elaborate simulations is reproduced, and the basic anatomy of
spiral structure generation in PPDs can
be understood in  terms of the response of sticky test particles at resonance.
I do not mean to imply that this is the only, or even
the primary, mechanism underlying the observed spirals in the
astrophysical environment, but it may be a fundamental driver of this
phenomenon.
\vskip 1\baselineskip
These spiral arms are
kinematic spirals;  they are not dynamic self-gravitating
structures, which would certainly be expected in cases where
the disk mass is a substantial fraction of that of the central star
(Dong et al. 2018).
Nonetheless, the fact remains that massive planets do form
in such disks and will, inevitably, have the effect of
generating spiral structure via this mechanism within a sufficiently
viscous surrounding gaseous disk.

\section{Response of gas disks to a weak bisymmetric perturbation.}

The equation of hydrodynamics in Lagrangian form, neglecting
pressure gradient forces and viscous stresses, is 
the equation of motion of a particle in the given gravitational potential.
That is to say, in the absence of thermal,
turbulent, viscous, or magnetic pressure forces, the motion
of an element of gas is described by an orbit, and in steady-state
gas stream lines correspond to simple non-intersecting
periodic orbits. But a stronger statement can be made:  Given
an ensemble of particles moving in a potential, filling the entire
volume of phase space permitted by the Hamiltonian constraint, and allowing
these particles to be ``sticky'' in the
sense that over some finite interaction distance the relative
radial
velocities of neighboring particles are reduced (in effect adding a viscous force that
resists compression of a fluid element), then we find that
such viscous dissipation forces the motion onto periodic
orbits; that is to say, periodic orbits arise as attractors in
the phase space of the system and each orbit family has its basin
Lake \& Norman 1984), the region of phase space within which a particle
will inevitably move toward the attractor.
\vskip 1\baselineskip
Only a subset of such periodic orbits can represent
streamlines:  those that do not cross other period orbits
(i.e., separate attractors with non-overlapping basins) and that
are not self-looping.
Otherwise hydrodynamical effects must intervene
and the nature of the gas flow is altered, but often in
way that is clearly related to the original periodic orbits.
This was demonstrated more than 40 years ago in Eulerian
hydrodynamical simulations which followed the gas response
in a galactic potential perturbed by a non-axisymmetric term ($cos(2\theta)$)
 (Sanders \& Huntley 1976). In these calculations the non-axisymmetric
forcing is
given a figure rotation such that all three principal resonances, 
inner-Lindblad (ILR), coronation, and  outer-Lindblad (OLR),  are present within
the numerical grid.  The assumed axisymmetric force law ($r^{-1.5}$) assures
that these resonances are equally spaced.
\vskip 1\baselineskip
The principal families of simple periodic orbits (most nearly
circular) near the inner resonance are elongated, but differ in
orientation by  90 degrees.  These two stable families
are designated X1 and X2 (Contopoulos \& Mertzanides 1977), where X1 is dominant
between the ILR and corotation and elongated parallel
to the major axis of the bar-like distortion;  X2 is elongated perpendicular to the
distortion and is dominant within the ILR or between
two ILRs.  For particle orbits the amplitude of the deviation from
circular motion increases near the resonance and the two perpendicular
families intersect, but
the hydrodynamical simulations demonstrate that the transition
from X2 to X1 is gradual and results in a
rotation of gas streamlines, which now crowd
in the locus of a trailing two-arm spiral.
\vskip 1\baselineskip
As a test I  
repeated this experiment using the Lagrangian
sticky particle technique employed here.
The details of this method have been described previously (Sanders 1998)
and are briefly
summarized here.  The particles, in two dimensions,
are disks all with radius $\sigma$.  A particle $i$ may be influenced
by a second particle $j$ at a separation $r_{ij}<\sigma$ where
that influence is weighted by a third-order polynomial
$w(x)$ ($x=r_{ij}/\sigma$) with coefficients chosen such that the peak is at
$w(0)=1.0$ and $w(x)$ falls smoothly to zero at $x=1$ ($w(x) = 1-3x^2+2x^3$)
At every time-step $\Delta t_k$ particle $i$ adjust its
velocity so as to reduce the velocity difference with
 each neighbor $j$, but only for approaching neighbors.  If
${\bf V}_{ij}$ is the component of the velocity difference
along the line
joining the two particles (${\bf V_{ij}}$ is a vector at the position
of $i$ pointing away from $j$),
then over this time step particle $i$ changes its
velocity by an amount given by the vector sum
$$\Delta {\bf v}_k=\alpha_k\sum_{j}^{r_{ij}<\sigma}{w(r_{ij}){\bf{V}}_{ij}}
\eqno(1)$$
with $$\alpha_k=\Delta t_k/t_s, \eqno(2)$$
where $t_s$ is a dissipation timescale typically taken to be  roughly a characteristic orbit timescale.

\begin{figure}
\begin{subfigure}[b]{0.6\textwidth}
\includegraphics[height=4cm]{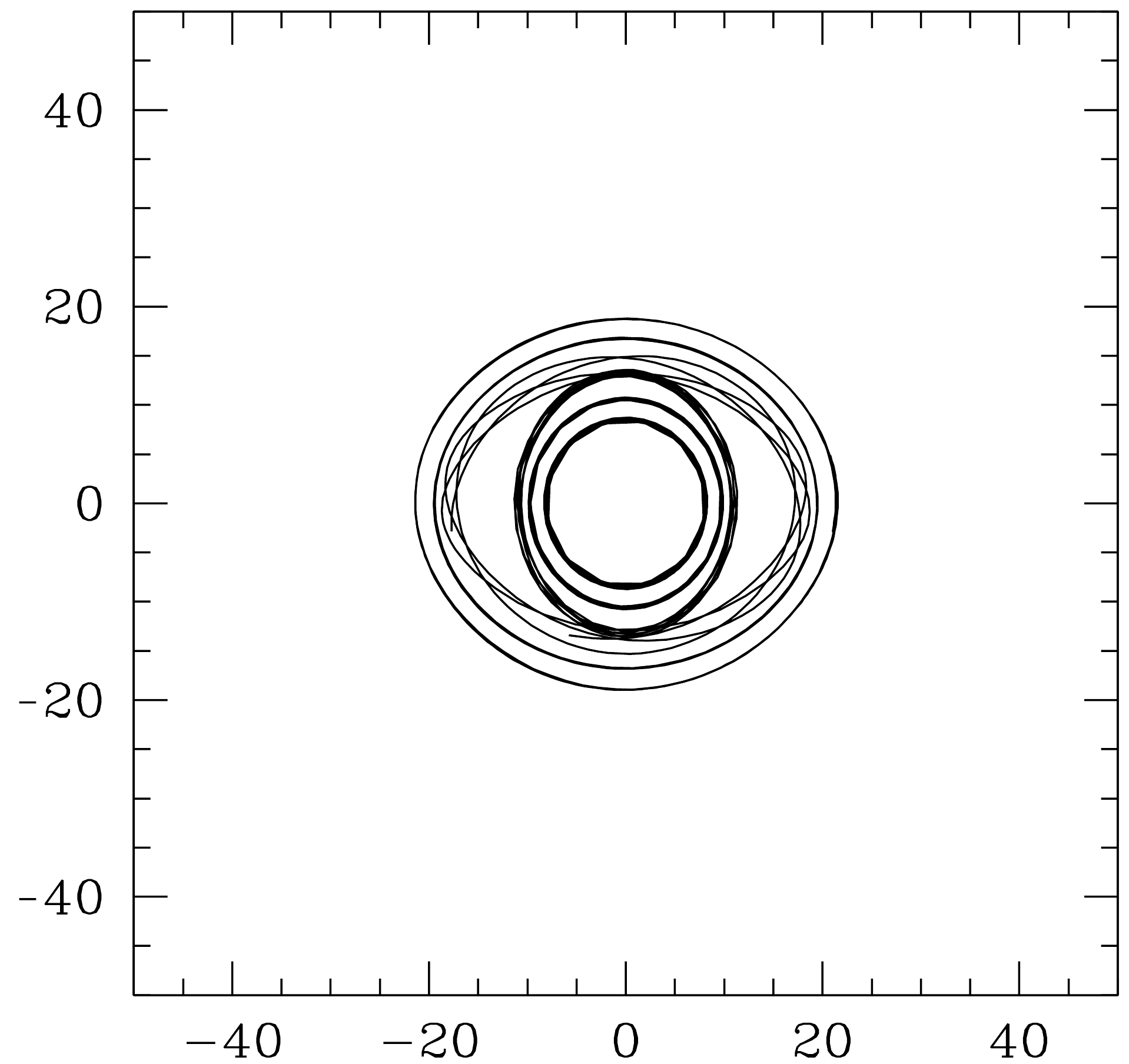}
\captionsetup{margin=10pt,font=small,labelfont=bf}
\caption{Periodic orbits in bar potential}
\end{subfigure}
\vspace{10mm}
\begin{subfigure}[b]{0.6\textwidth}
\includegraphics[height=4cm]{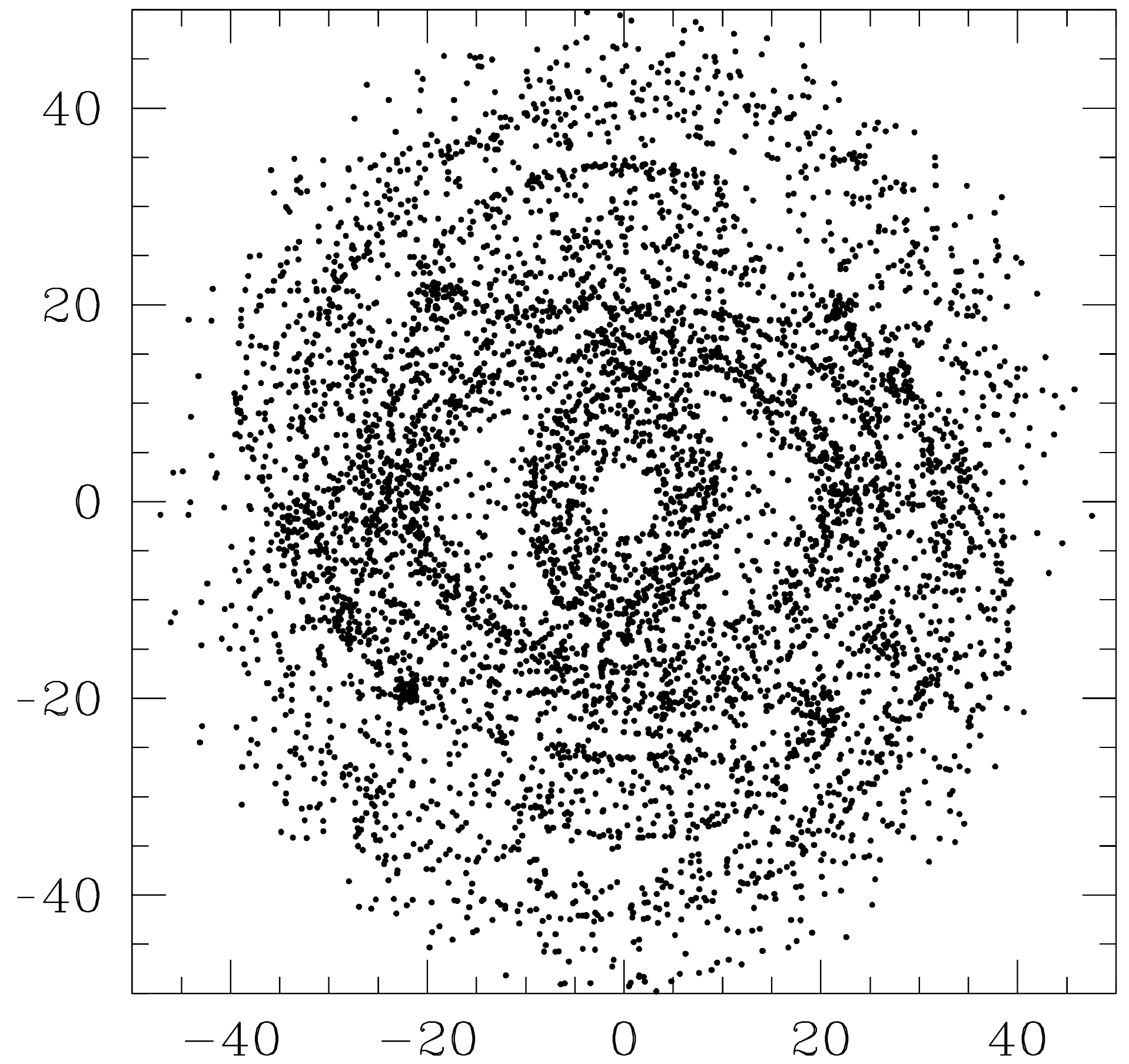}
\captionsetup{margin=10pt,font=small,labelfont=bf}
\caption{Particle ensemble in bar potential}
\end{subfigure}
\begin{subfigure}[b]{0.6\textwidth}
\includegraphics[height=4cm]{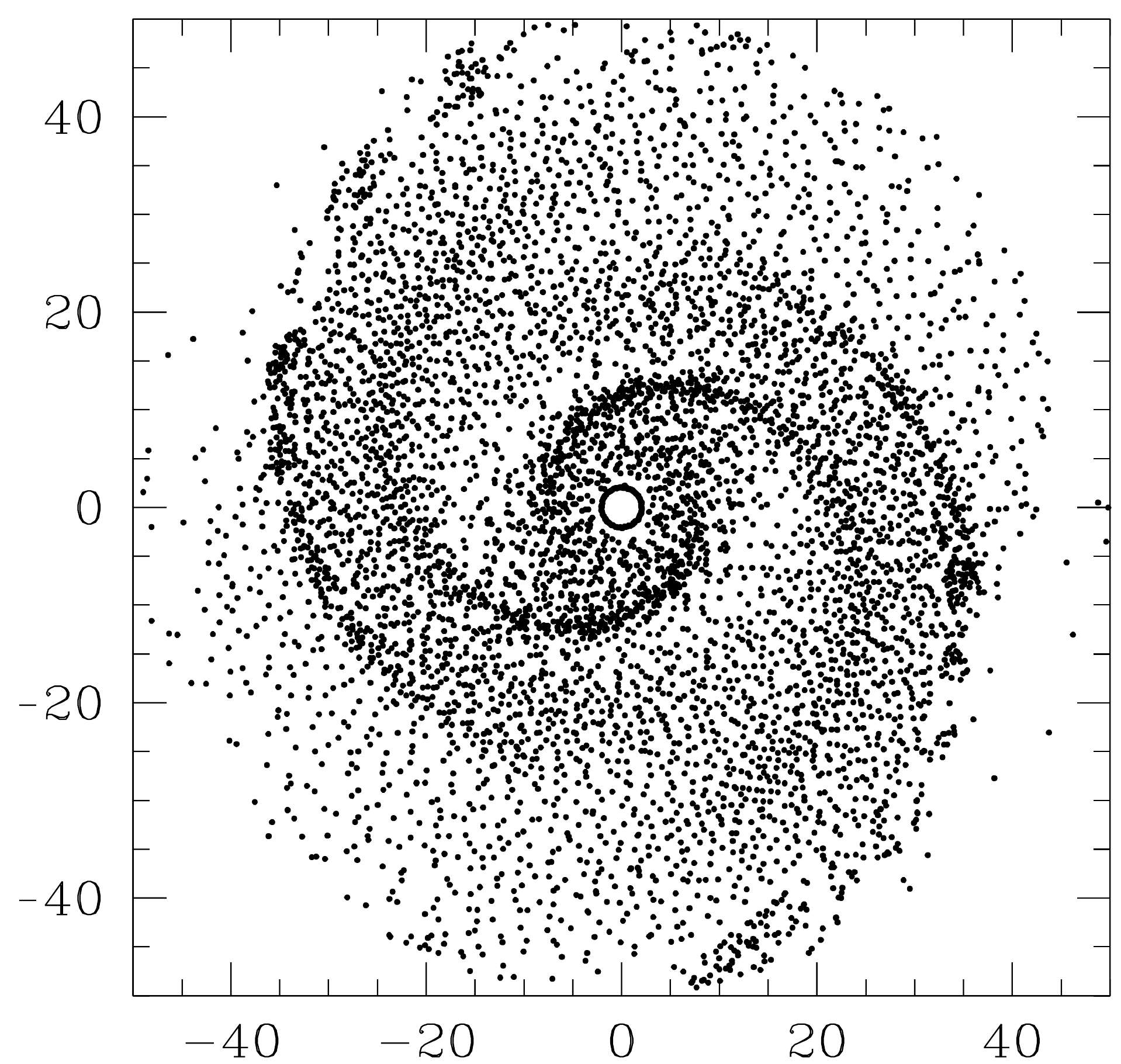}
\captionsetup{margin=10pt,font=small,labelfont=bf}
\caption{Ensemble in bar potential with viscosity}
\end{subfigure}
\vspace{10mm}
\begin{subfigure}[b]{0.6\textwidth}
\includegraphics[height=4cm]{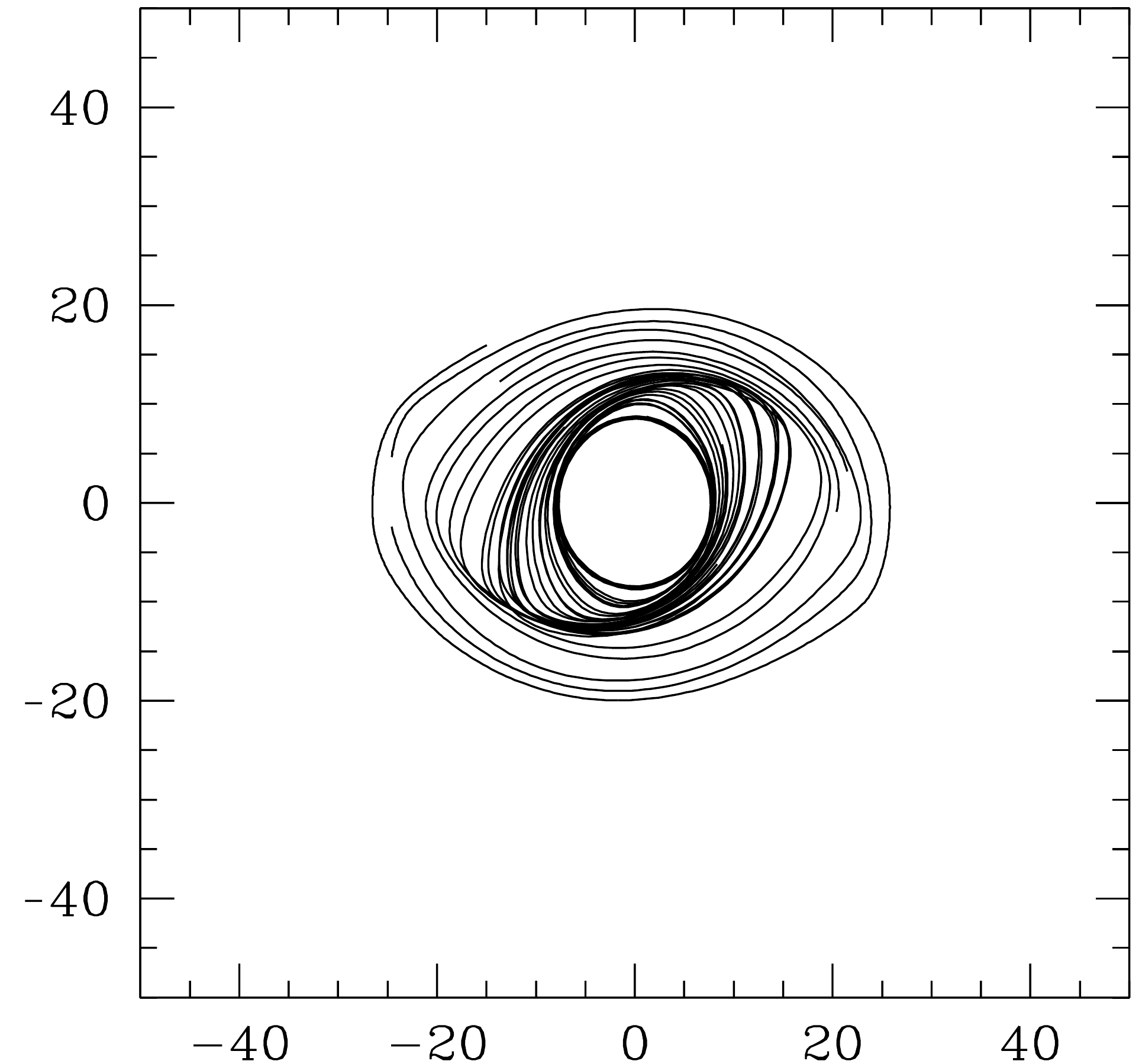}
\captionsetup{margin=10pt,font=small,labelfont=bf}
\caption{Gas orbits near resonance in full simulation}
\end{subfigure}
\caption{Particles and gas in weak rotating bar potential}
\end{figure}

This method provides an explicit kinematic
bulk viscosity that is proportional
to the local velocity divergence, but only if that divergence is
negative (converging flow).  The method manifestly conserves
linear momentum (the effect of particle $j$ on particle $i$ is equal but
opposite that of particle $i$ on particle $j$);  angular momentum is also
conserved to high precision in rotating viscous fluid in
an axisymmetric potential.  There is no explicit shear
viscosity in this
scheme, but because it is dissipational the method does not conserve
energy.  Consequently, there is some radial re-arrangement of
the density of particles in an axisymmetric rotating fluid, but this
happens slowly,  on a  timescale of 50 characteristic rotation periods.
\vskip 1\baselineskip

In repeating the Sanders-Huntley experiment I took 6000 finite size particles
initially uniformly covering a circular area with a radius
of 40 distance units (one unit may be taken as 10 pc).  For each particle
$\sigma = 3.5$ units so that on average one particle   overlaps 45  neighboring
particles.  The strength of the non-axisymmetric perturbation is initially
zero, but grows to a maximum over one pattern rotation period corresponding
to an azimuthal force of 0.05 of the axisymmetric force.  The pattern
speed is such that the three principal resonances are located at radii
15, 30, and 45.  The setup and the results are shown in Figure 1.

\vskip 1\baselineskip
Figure 1a shows the principal periodic orbits in the perturbed potential
near the inner resonance:
the X1 family elongated
parallel to the major axis of the bar distortion (along the X-axis),
and the X2 family perpendicular
to the bar distortion.  These two families intersect
near the resonance, and therefore cannot represent gas streamlines.
Figure 1b illustrates the distribution of 6000 particles after five
pattern rotation periods in
the fully perturbed potential, but with no dissipation (this
is a pure orbit calculation). We see the pattern of periodic
orbits in the crowding of particles which gives rise to
rings and gaps.   Figure 1c
is the distribution of 6000 dissipative particles in the complete
non-axisymmetric potential after five bar rotation periods with the
viscous force (Eq. 1) in place.
The trailing spiral structure is present throughout the radial range
of the simulation
and is similar (but not identical) to that found by Sanders-Huntley
in the Eulerian scheme
on a 40X40 cartesian grid;  in the earlier simulation there is even a
trace of the double
spiral structure seen here. Figure 1d shows the paths of several particles near
resonance taken from the full simulation (Fig. 1c).
These are  the gas streamlines; the gradual rotation
of the streamlines by 90 degrees between X2 and X1 is evident
and results in the trailing spiral pattern.

Overall, the method applied here
provides an acceptable solution for the gas distribution and flow
in the perturbed potential in the highly supersonic limit.

\section{Gas flow in a protoplanetary disk perturbed by a giant planet}

The proposal is that the generation of spiral structure in a PPD
containing a Jupiter-mass  planet
is fundamentally the same as in the perturbed galaxy disk considered
above with respect to the role of periodic orbits.
The origin of the
inverse square force is at the center, and the source can
be scaled to one solar mass.   The planet is on a circular orbit at a
radius of 32 units where one unit can be scaled to 1 AU.
In the Keplerian potential, this places the ILR and OLR at
20 and 42 AU, respectively.   Combined with
a velocity unit of 1 km/s, the distance scaling
of 1 AU and a mass scale of one solar mass, the time unit is $4.75$ years. 
Scaling is straightforward
in this gravitational central force problem with no additional physics;
keeping $GM$ fixed while scaling distance by a factor $f$ is equivalent to
scaling the time unit by a factor of $f^{1.5}$.

\vskip 1\baselineskip
Figure  2 illustrates the structure of several periodic orbits in the
rotating frame of the planetary orbit near the
ILR. The different panels correspond to four different
values of the planetary mass: 0,001, 0.002,
0.004 and 0.006 in units of the central mass (1, 2, 4, and 6
Jupiter masses or $M_J$).
The two perpendicular orbit families are evident in the perturbed Keplerian
potential.  In Fig. 2a (one Jupiter mass) these periodic orbits are
seen to barely overlap,  but for 2, 4, and 6 $M_J$ the overlapping
becomes more pronounced, as do deviations from m=2 symmetry.
From the above arguments this would imply that the spiral structure should
be weak for the 1 $M_J$ perturber, but increasingly
conspicuous for the higher mass perturber, and indeed this
is evident in the simulations including dissipation.
\vskip 1\baselineskip
Figure 3 illustrates the response of the gas disk corresponding to these
same planetary masses in the rotating frame. Again
I take 6000 particles initially distributed with constant density in
a disk with a radius of 40 AU.
For every particle $\sigma = 1.8$, so initially the average
number of interacting neighbors is 12.  To avoid high
accelerations an inner boundary is taken at 5 AU; particles
crossing the inner boundary are reflected.  The panels are snapshots showing the
distribution of the ensemble of dissipating particles after
four planet orbital periods (780 years).
In all cases, a quasi-steady state is reached after
two rotation periods.  To avoid shocking the system
I have also run cases in which the planet mass builds
up slowly, over one revolution period, but this makes no difference in the final gas
distribution,

\begin{figure}
\begin{subfigure}[b]{0.6\textwidth}
\includegraphics[height=4cm]{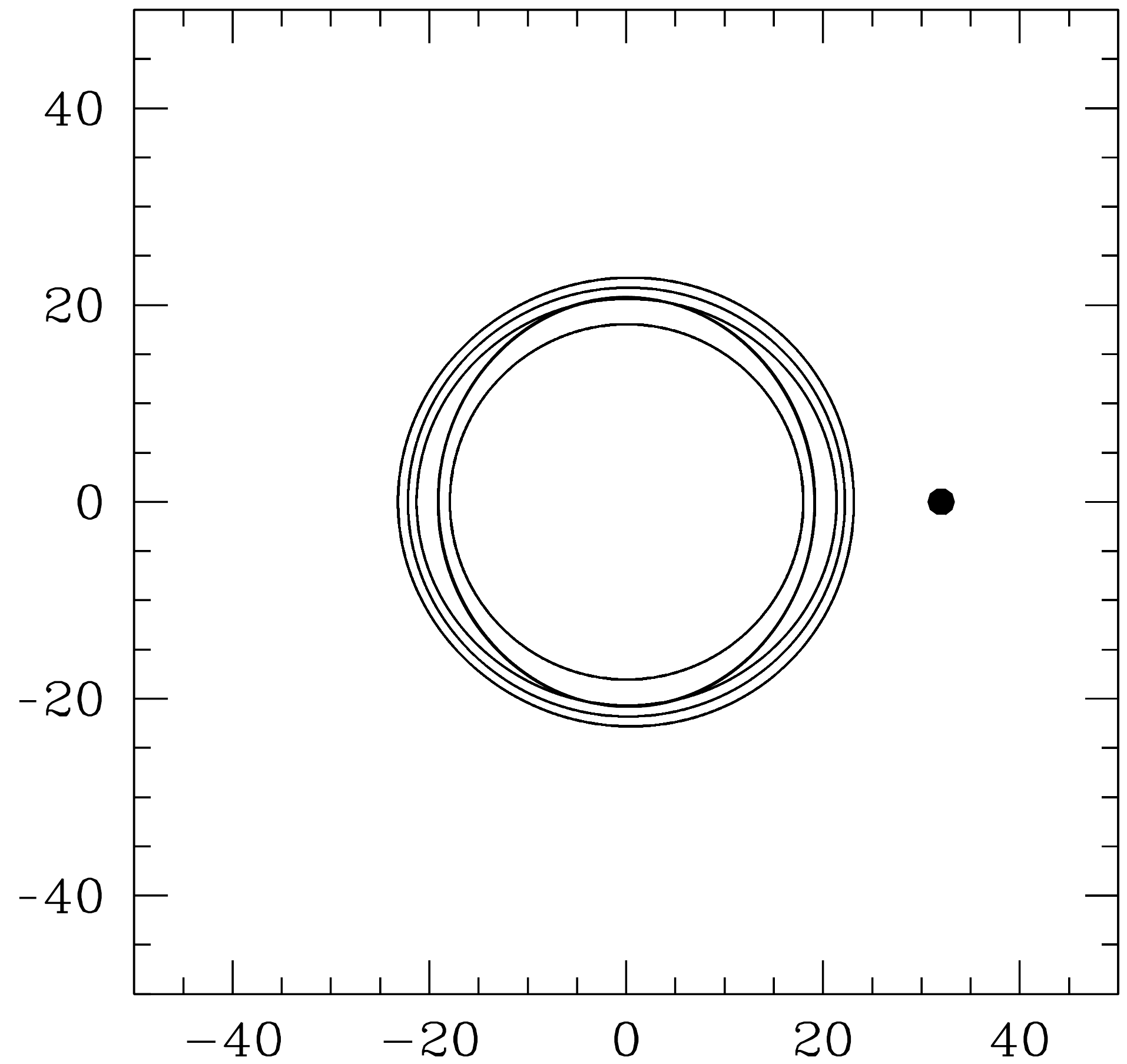}
\captionsetup{margin=10pt,font=small,labelfont=bf}
\caption{$M_p = 1\, M_J$}
\end{subfigure}
\vspace{10mm}
\begin{subfigure}[b]{0.6\textwidth}
\includegraphics[height=4cm]{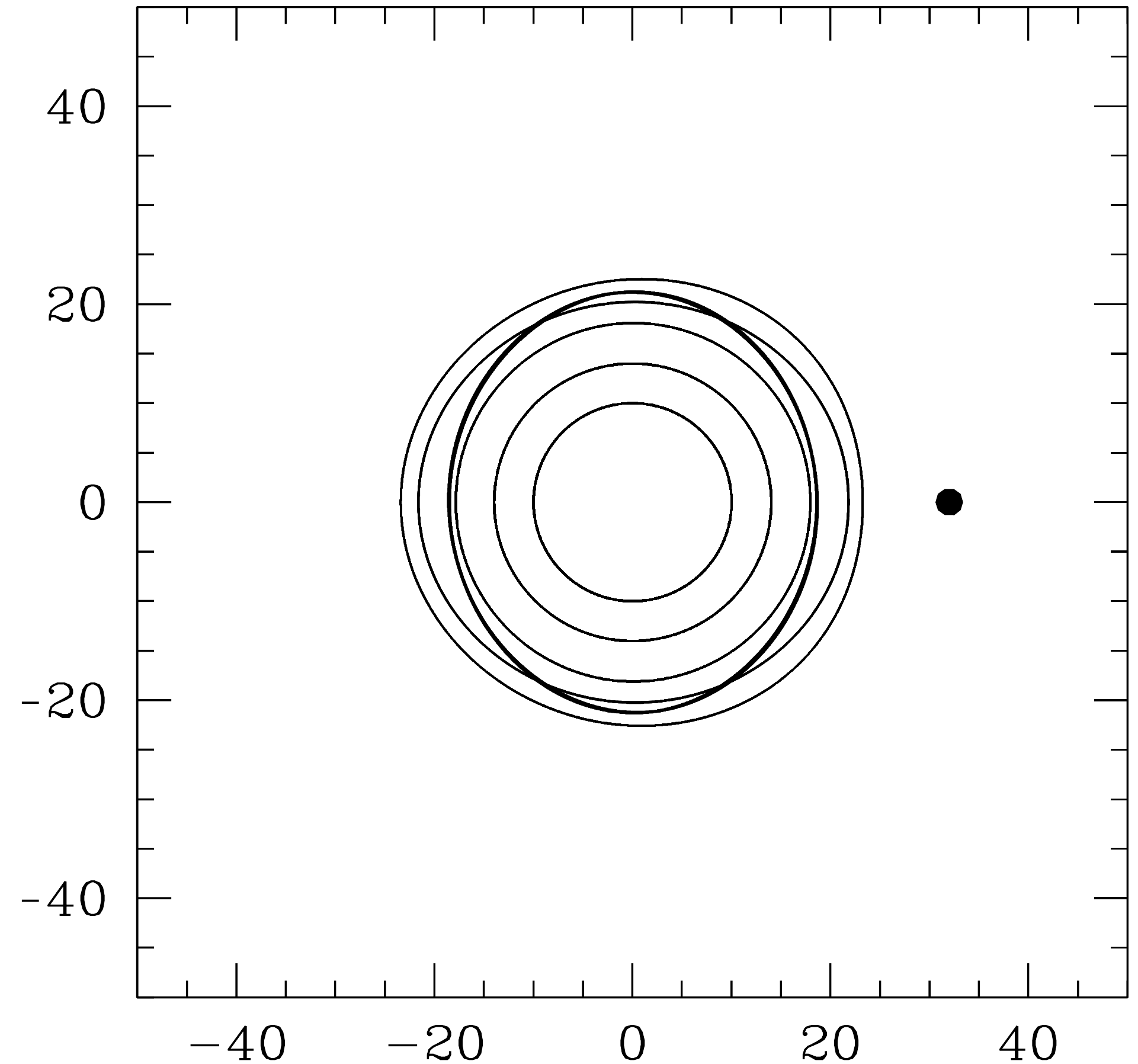}
\captionsetup{margin=10pt,font=small,labelfont=bf}
\caption{$M_p=2\, M_J$}
\end{subfigure}
\begin{subfigure}[b]{0.6\textwidth}
\includegraphics[height=4cm]{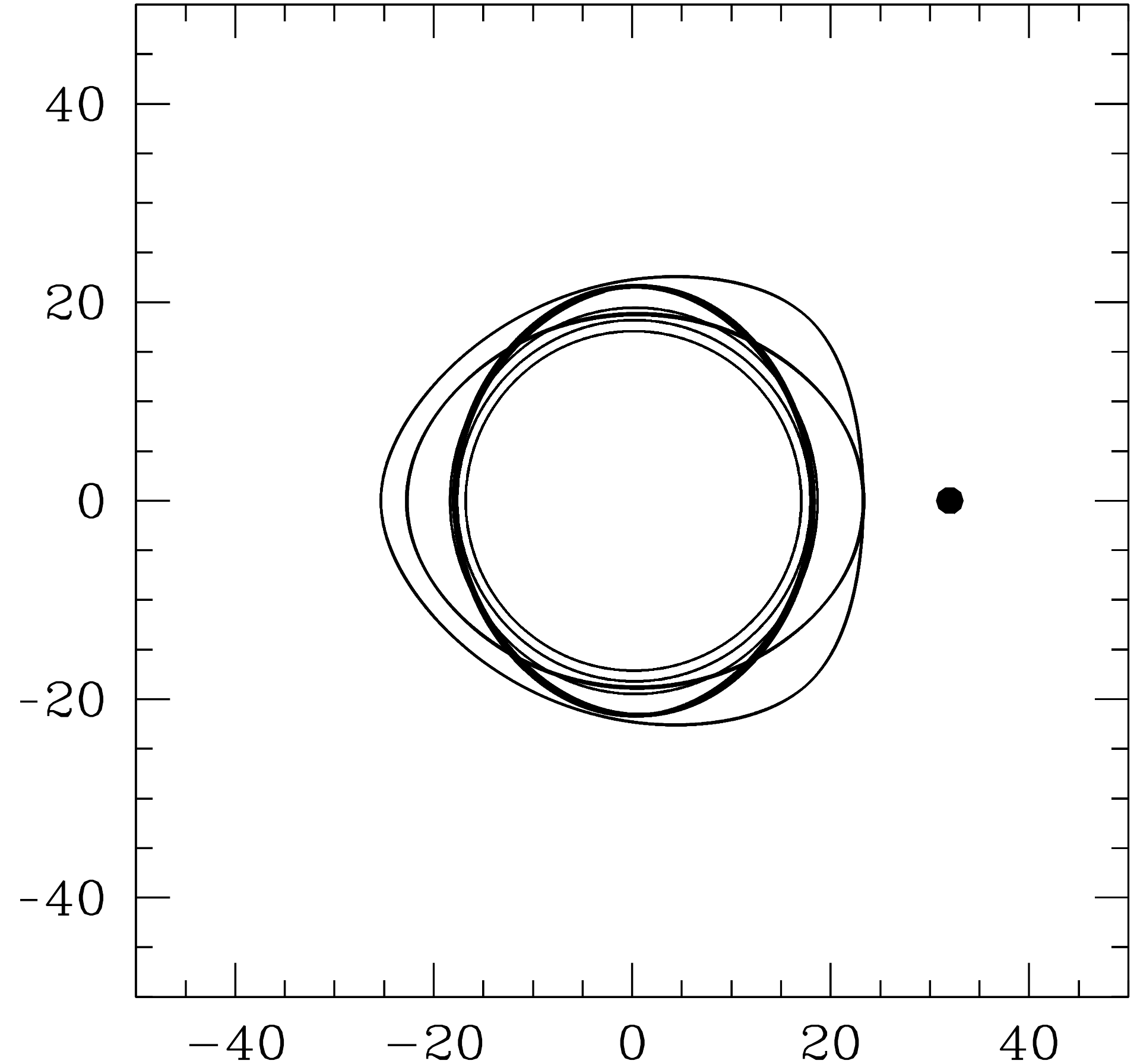}
\captionsetup{margin=10pt,font=small,labelfont=bf}
\caption{$M_p=4\, M_J$}
\end{subfigure}
\vspace{10mm}
\begin{subfigure}[b]{0.6\textwidth}
\includegraphics[height=4cm]{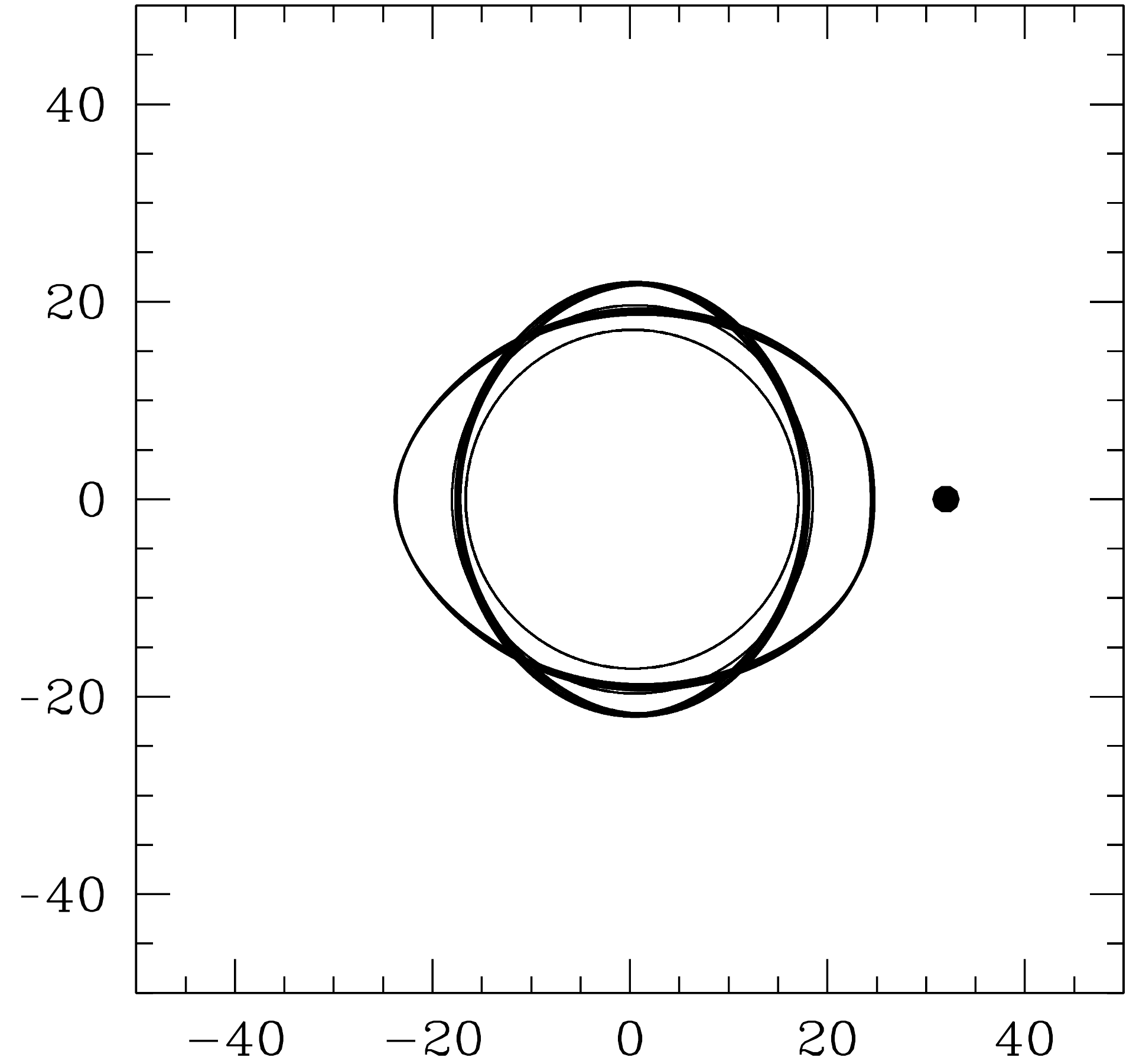}
\captionsetup{margin=10pt,font=small,labelfont=bf}
\caption{$M_p=6\, M_J$}
\end{subfigure}
\caption{Periodic orbits in the rotating frame near inner resonance with
various planet masses.  Corotation is at $r=32$ and the inner resonance at $r=20$.}
\end{figure}

\begin{figure}
\begin{subfigure}[b]{0.6\textwidth}
\includegraphics[height=4cm]{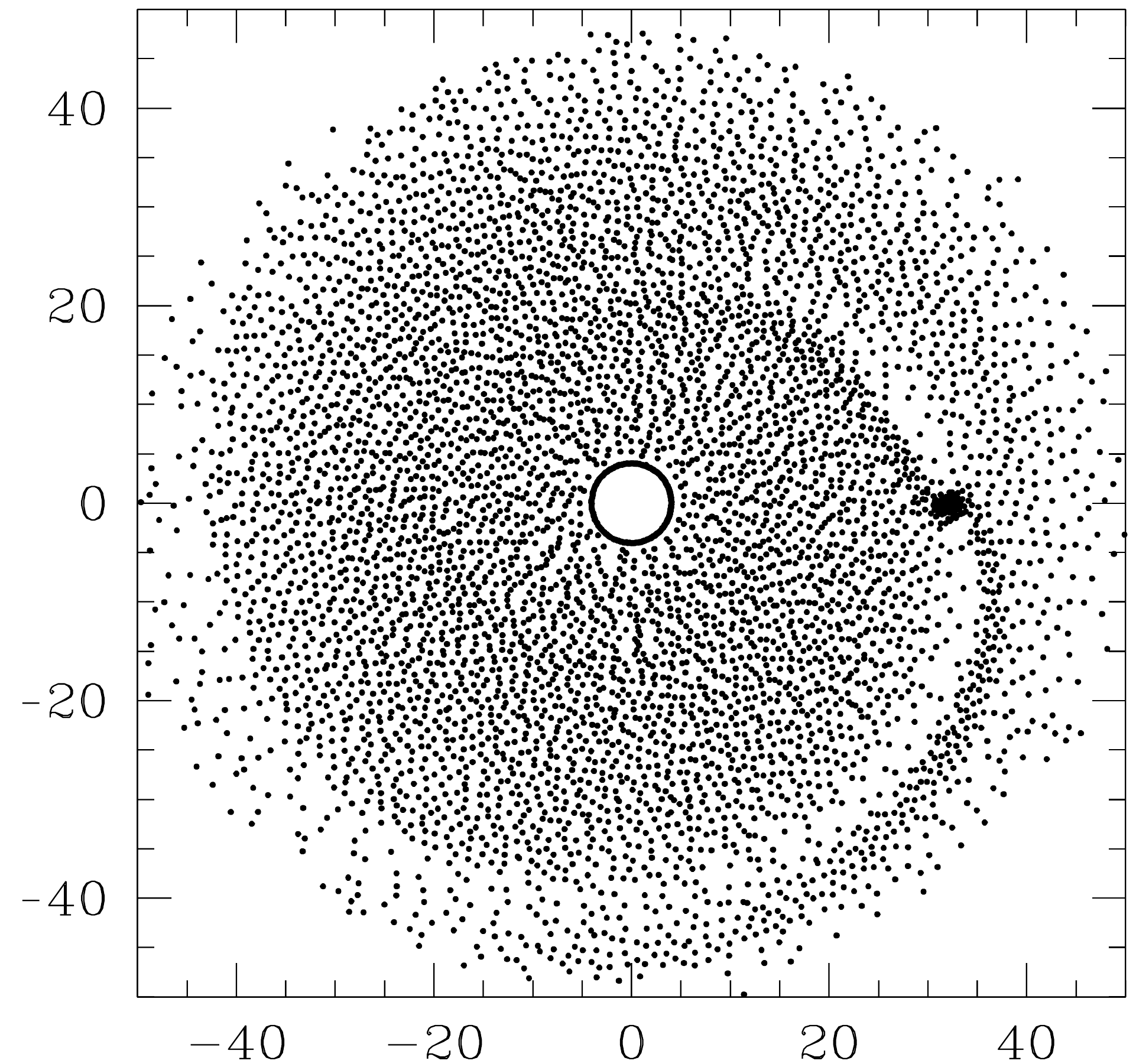}
\captionsetup{margin=10pt,font=small,labelfont=bf}
\caption{$M_p = 1\, M_J$}
\end{subfigure}
\vspace{10mm}
\begin{subfigure}[b]{0.6\textwidth}
\includegraphics[height=4cm]{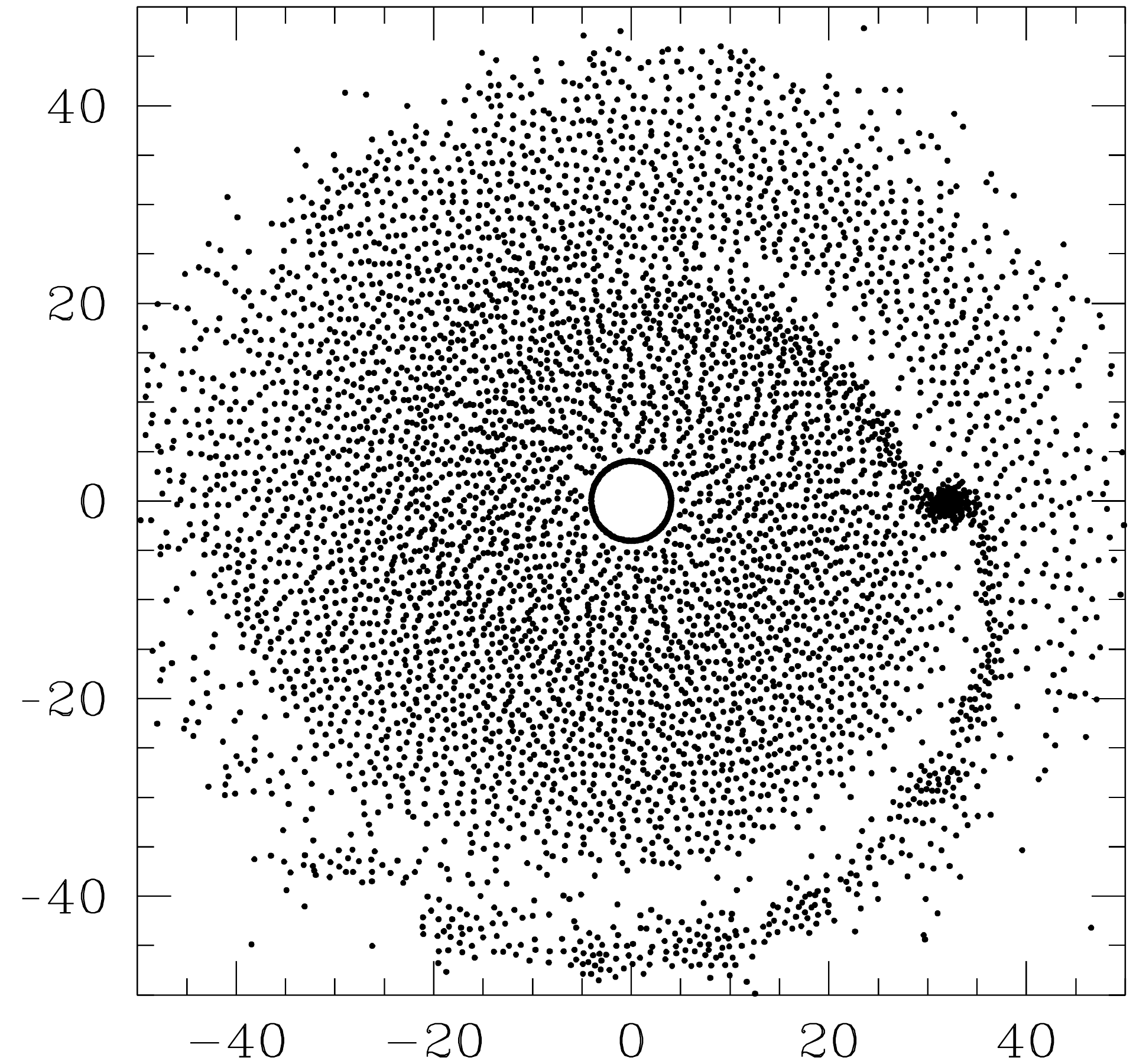}
\captionsetup{margin=10pt,font=small,labelfont=bf}
\caption{$M_p=2\, M_J$}
\end{subfigure}
\begin{subfigure}[b]{0.6\textwidth}
\includegraphics[height=4cm]{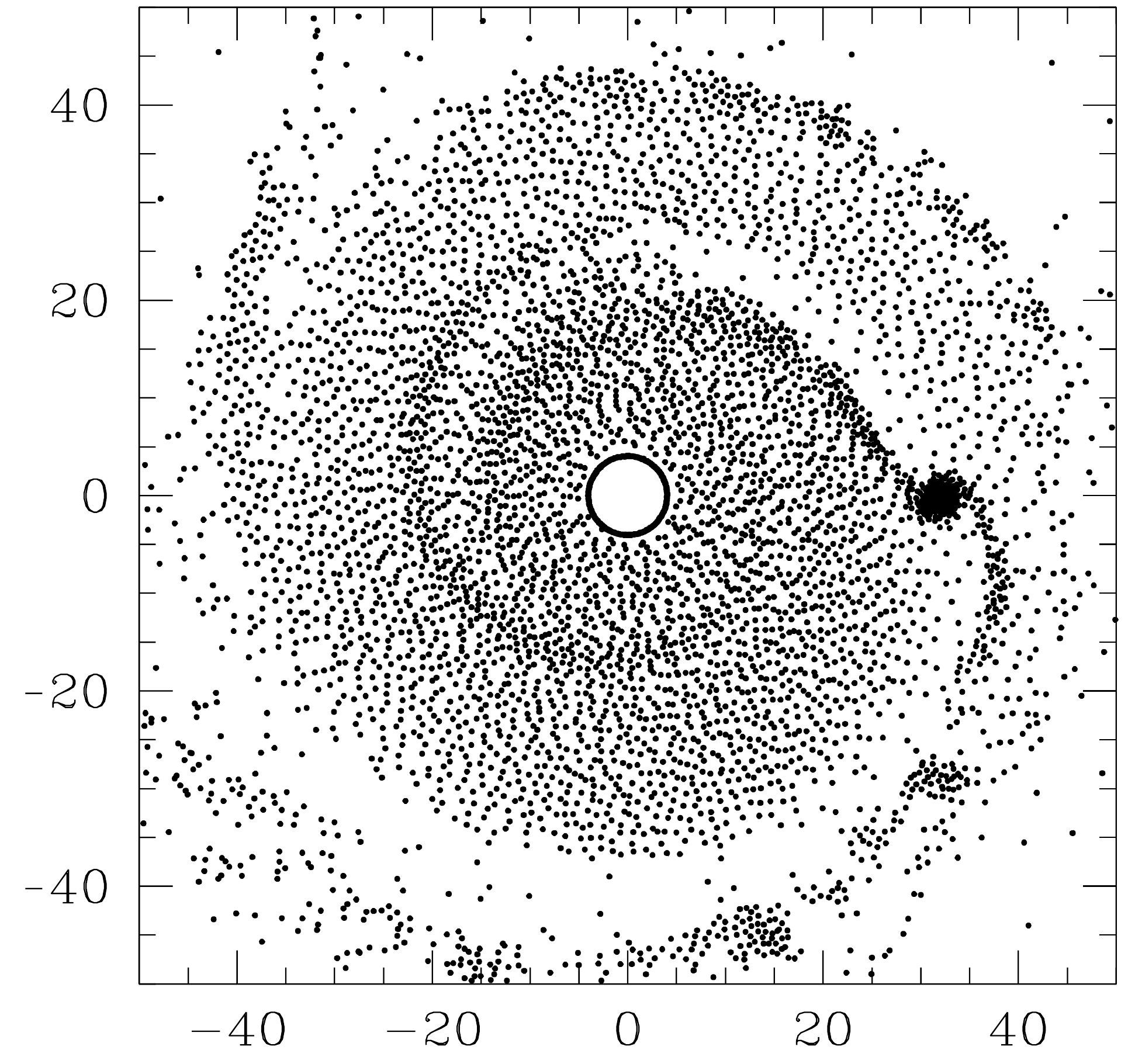}
\captionsetup{margin=10pt,font=small,labelfont=bf}
\caption{$M_p=4\, M_J$}
\end{subfigure}
\vspace{10mm}
\begin{subfigure}[b]{0.6\textwidth}
\includegraphics[height=4cm]{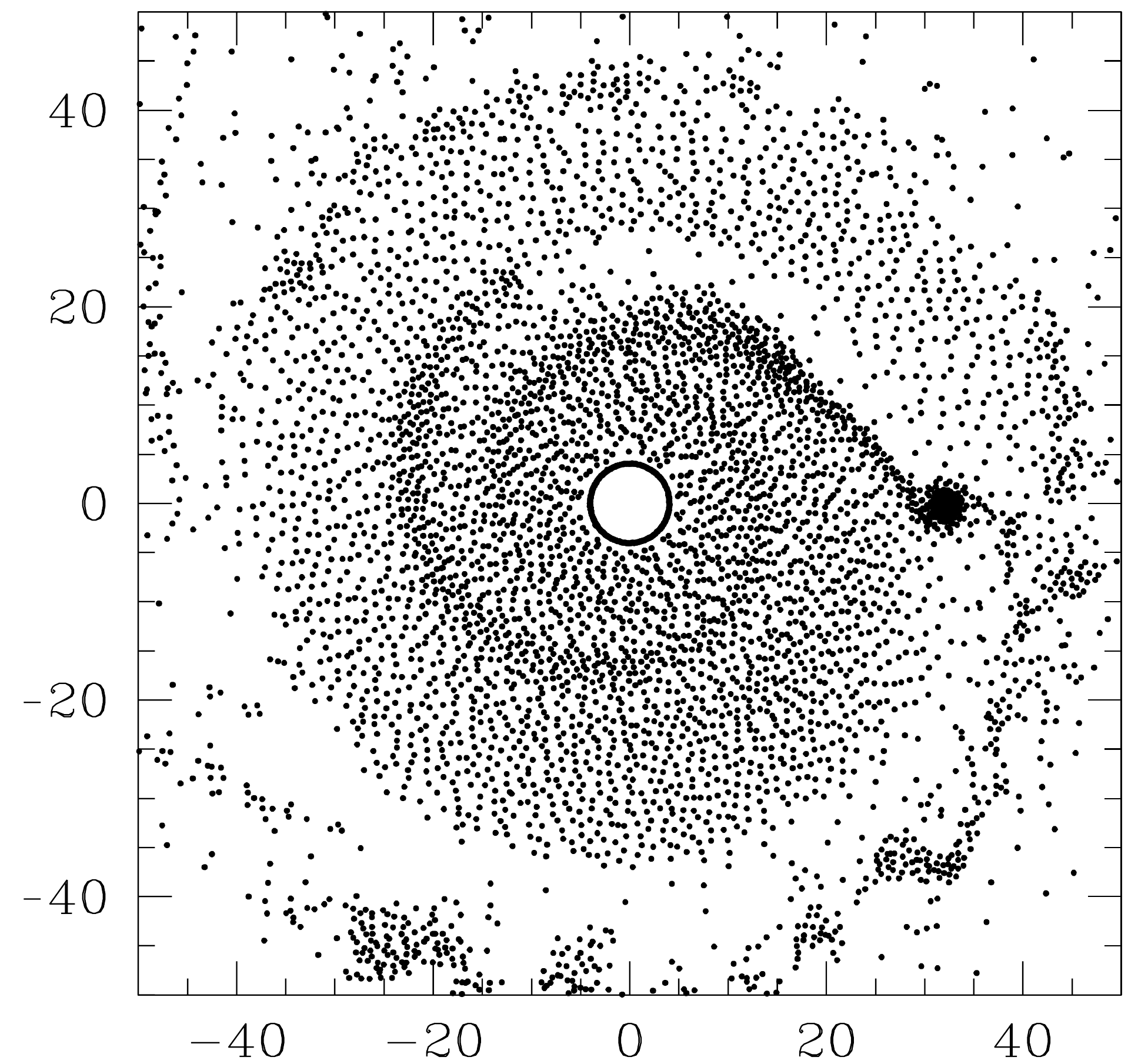}
\captionsetup{margin=10pt,font=small,labelfont=bf}
\caption{$M_p=6\, M_J$}
\end{subfigure}
\caption{Snapshots of the quasi-steady-state density distribution after 4.5 planet rotation periods in the fully dissipative gas of 6000 particles.}
\end{figure}

In the case of a one
Jupiter mass perturber, a faint spiral structure is
visible, as would be expected from the slight overlapping
of periodic orbits, but as the mass of the perturber increases,
the spiral response becomes stronger.
There are two principal arms that wrap through roughly 200 degrees.
The breakdown from bisymmetry due to the planet is
especially evident in the
primary arm that connects to the planet at corotation. There is a cluster
of particles about the planet in all cases; this  corresponds to
the Hill sphere within which particles are permanently trapped by
the planet (radius of about 3.5 AU for the 4 $M_J$ case).

\vskip 1\baselineskip
The amplitude
of the spiral response appears to be roughly proportional to the
mass of the perturber over this range of planetary masses.
This is illustrated in Fig.\ 4, which is a plot of the
gas surface density (smearing particle mass by the smoothing function
$w(x)$ over $\sigma$) as a function of azimuthal angle at a radius of 20 AU
for a planet masses of 2 $M_J$ (solid curve) and 6 $M_J$ (dashed curve).
The amplitude of the response increases
with the mass of the perturber, as does the asymmetry between
the two principle arms.  This
is expected because of the increasing m=1 signal with
planet mass in the perturbed potential.
This figure also illustrates an additional
marker of the basic spiral structure:  the angular separation between
the primary and secondary arms.  At 20 AU ($r/r_p = 0.62$)
this varies from 140 degrees (2 $M_J$) to 165 degrees (6 $M_J$), which is
broadly consistent with the results given by Bae \& Zhu (2018b).
\vskip 1\baselineskip

\begin{figure}
\includegraphics[height=7cm]{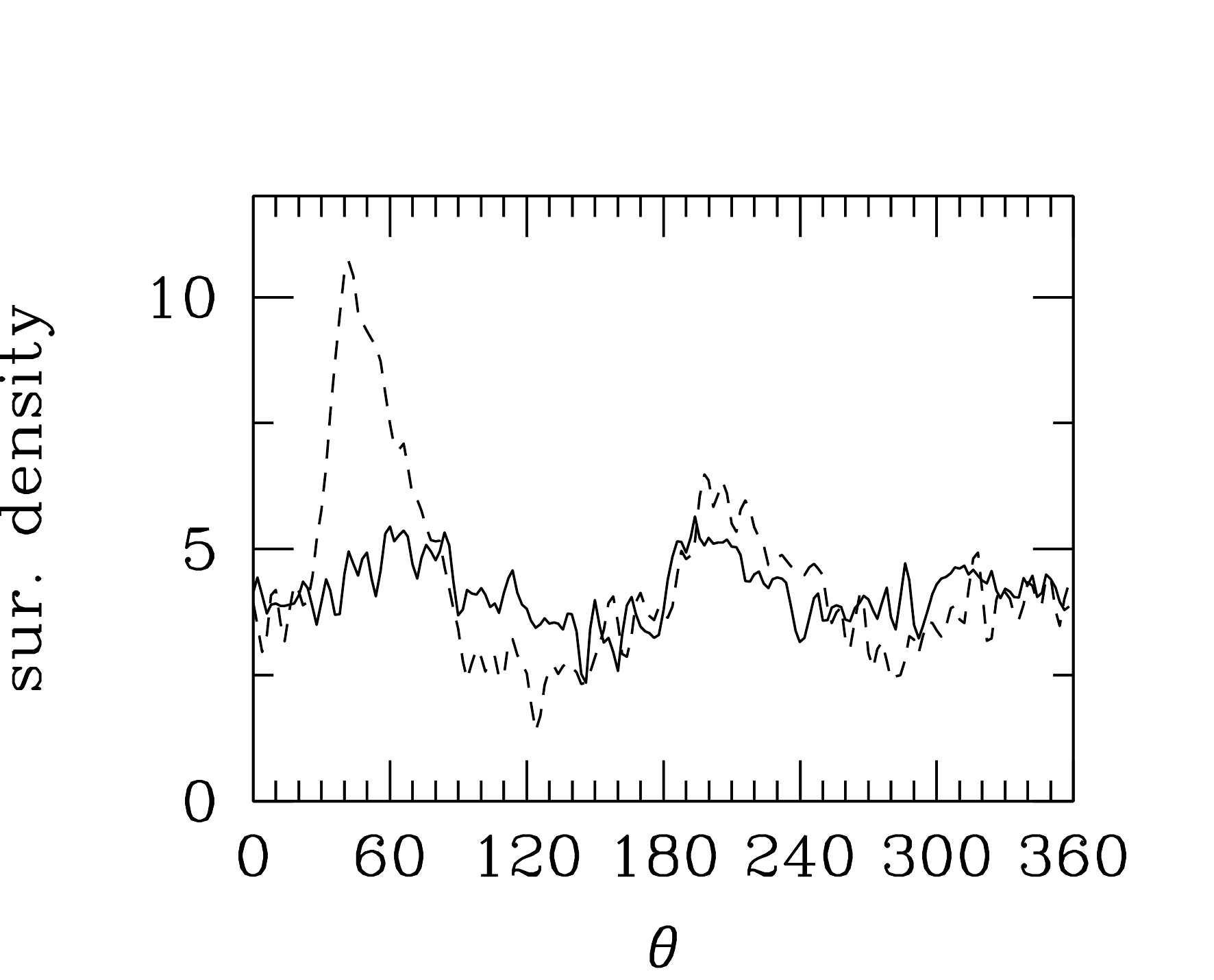}
\captionsetup{margin=10pt,font=small,labelfont=bf}
\caption{Azimuthal distribution of gas surface density for 2 $M_J$ (solid curve) and
6 $M_J$ (dashed curve).}
\end{figure}

\begin{figure}
\includegraphics[height=9cm]{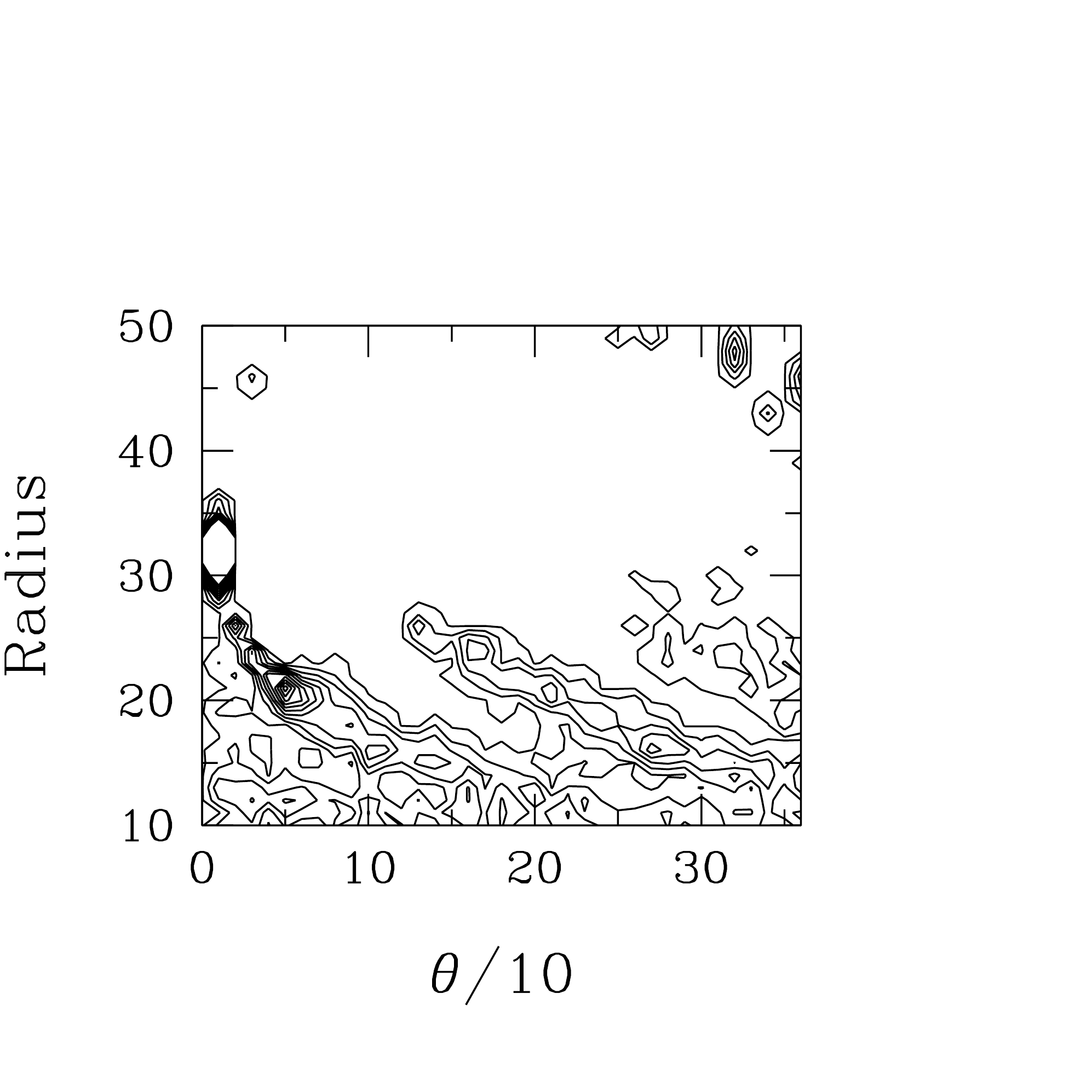}
\captionsetup{margin=10pt,font=small,labelfont=bf}
\caption{Gas density distribution for the 6 $M_J$ case in  polar projection
corresponding to panel 3d.}
\end{figure}

\begin{figure}
\includegraphics[height=6cm]{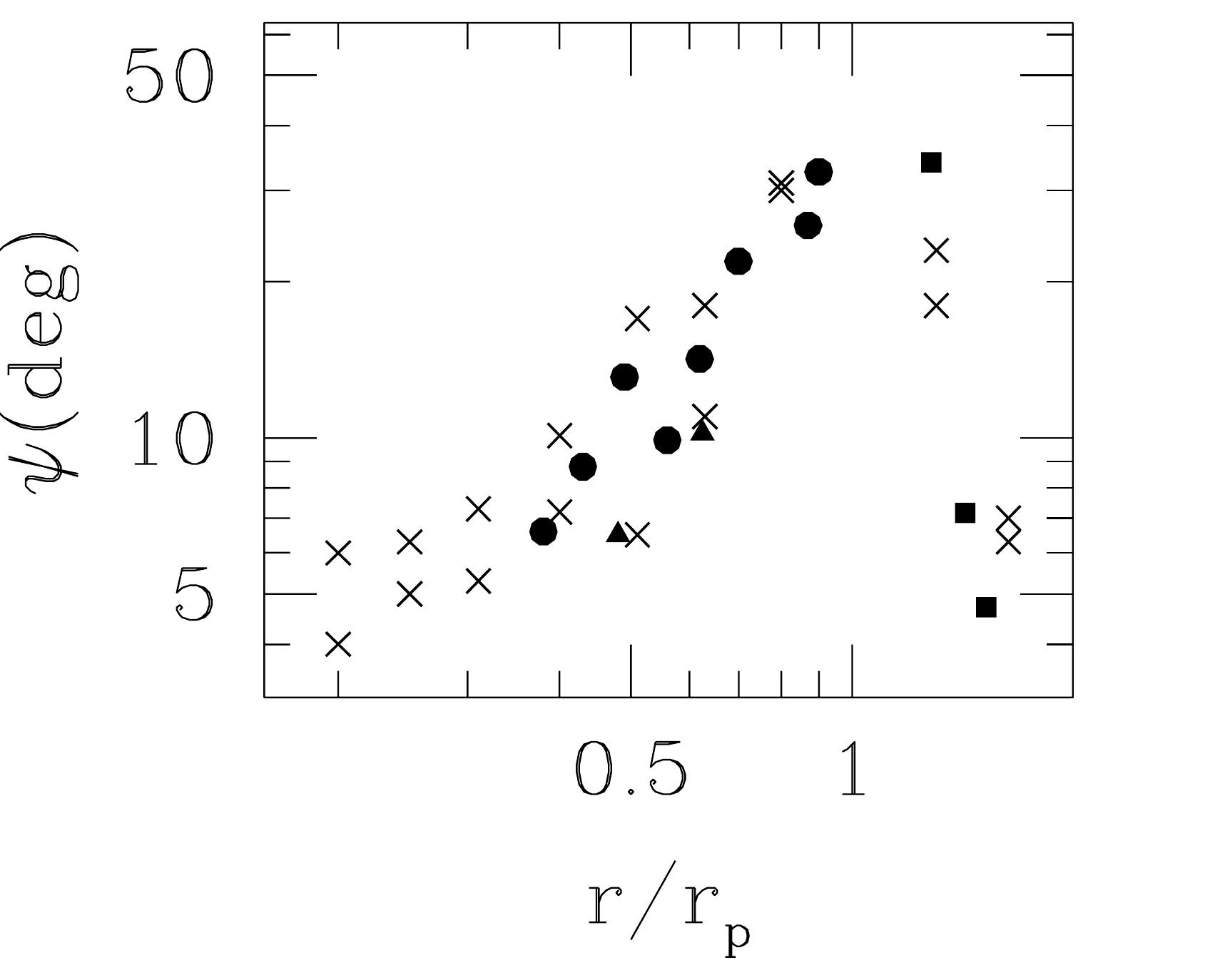}
\captionsetup{margin=10pt,font=small,labelfont=bf}
\caption{Pitch angles as a function of radius scaled to planet radius
for 6 $M_J$:  
 circles and squares are the data points from the present calculations (triangles are secondary
spirals);  crosses are the data points from the simulations of Zhu et al. (2016).}
\end{figure}

For both
principal arms the  pitch angle decreases moving inward from corotation;
thus, the geometry of the arms is not
 described well by a logarithmic spiral.  To estimate the run of
pitch angle with radius it is best to consider the density distribution
displayed in a polar map, as in Fig. 5 for the $M_p = 6\, M_J$ case
(density contours shown in the
$\theta$-$r$ space). The two principal arms (and a faint third arm)
are evident in the figure,  as
is  the deviation, most conspicuous for the primary arm from a
constant pitch angle.
\vskip 1\baselineskip
With this display one can determine the run of pitch angle via the relation
$\psi=d{ln(r)}/{d\theta}$; the results are shown in Fig. 6 where
the run of pitch angle is plotted against radius in terms of the semi-major
axis of the 6 $M_J$ planet.  The results for the primary amd secondary
inner arms are shown, as well as the outer arm,
and are compared to the numerical simulations of
Zhu, et al. (2015). 
The form and range
of the pitch angle dependence do not vary greatly with planetary mass,
reflecting the fact that streamline rotation of 90 degrees
is the maximum possible over the range between ILR and corotation
independent of planetary mass.  It is also clear that the arms tend to
wrap into a ring at about 0.4 $r_p$, that is, within the inner resonance at 0.62 $r_p$.
In the higher resolution calculations of Bae \& Zhu 2018b, the spirals continue to
wrap within   this radius.

These results on pitch angle, asymmetry, and
arm separation are generally consistent
with the sophisticated simulations of Zhu et al.  This is significant
because the only physics in the present calculation is orbital motion
in the perturbed Newtonian potential with a crude algorithm for
viscosity.

\section{Role of dissipation}

It should be noted that  
the mechanism described here for the role of viscous dissipation  differs from most previous analyses of spiral
structure generated in gaseous discs via interaction with planets.
The seminal work on this subject is 
Goldreich \& Tremaine (1979):
in the linear limit spiral wakes in a gaseous disk perturbed by
an orbiting planet are explained as a result of
constructive interference among a set a wave modes propagating at the sound
speed and having different azimuthal
wave numbers.  These linear calculations are inviscid, but include self-gravity
of the gaseous disk.   A spiral wave
is formed from the inner Lindblad resonance to corotation, and Goldreich and
Tremaine emphasize that in the inviscid case no spiral waves propagate
without some degree of self-gravity.

In the present simulations there is no self-gravity of the gas, but it is argued
that the large-scale spiral structure
results from
trapping of gas on periodic orbits;  near resonance these orbits intersect
and dissipation intervenes to promote the gradual rotation of streamlines
at resonance which crowd into a two-arm spiral pattern.  Viscous dissipation
is the only gas dynamical effect, and here it plays a critical role in the formation of
coherent structure.  The question arises:  How do these two descriptions relate
to one another: are they separate explanations or the same mechanism?

\onecolumn
\begin{figure}
\includegraphics[height=12cm]{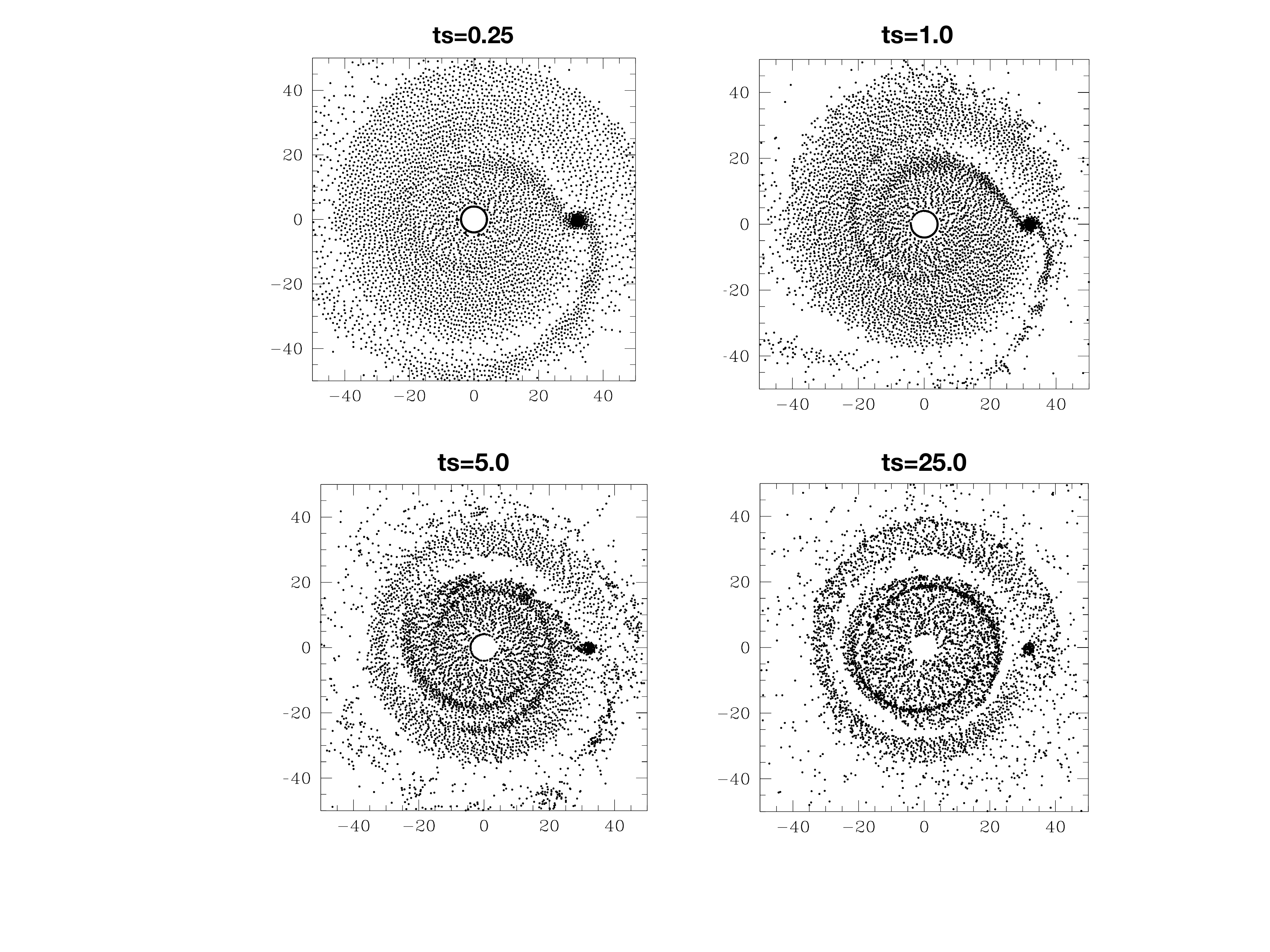}
\captionsetup{margin=10pt,font=small,labelfont=bf}
\caption{Effect of the dissipation timescale on the  morphology of structure.  The
sequence a-d is in the sense of decreasing viscous dissipation.
}
\end{figure}

In the simulations the dissipation is promoted by an artificial bulk
viscosity that inhibits the crossing of streamlines.  For a disk
initially in pure rotational motion, the finite size particles jostle
one another due to the differential rotation,
and a random velocity appears that initially grows until the energy
loss rate becomes equal to the energy gain.  In the simulations with
$\sigma = 1.8$ and a viscous timescale of 1 au/kms$^{-1}$ this equilibrium
velocity is $\Delta V \approx 0.03 $ km/s at $r_p=32$.
This could be taken as the effective signal speed in the system; a disturbance
is propagated from the center to corotation on a timescale $r_p/\Delta V$
or roughly 30 planetary revolution periods.  But in the simulations the
structure develops on the much shorter dynamical (orbit) timescale, which is relevant to the
trapping on periodic orbits.

It can be shown that for this algorithm (Eq.\ 1) the resultant kinematic
(bulk) viscosity in units of $r_pV_p$ at the   radius of the planet is given by $\nu\approx f(\sigma/r_p)^3
{t_s}^{-1}$, where $f$ is a factor (1/8 to 1/16) correcting $\sigma$ to
an effective particle size, and $t_s$ is the dissipation timescale in the adopted
units (1 au/km/s).
This amounts to $2.5 \times 10^{-5} \, (f/0.5)^3$ at $r_p$.

Previous discussions of viscosity in this context (Kley 1999; Bryden et al. 2000) have
 primarily focused on the longer timescale
problem of accretion onto the forming planet rather than the role
with respect to structure formation.  Kley in particular considered separately
an artificial kinematic bulk viscosity as well as shear, and his scaled
viscosity coefficient was approximately that derived above. It is interesting
that the spiral structure that developed was also similar
to that described here.
Like several cases considered by Kley, the present simulations do not contain
an explicit shear viscosity;  only the radially approaching components
of the relative velocity of finite size gas particles are considered, so it
is not meaningful to discuss these simulations in terms of the $\alpha$ disk
assumption.  Explicit shear viscosity could have been included in eq. 1,
but this was not relevant because the essential role of viscous dissipation
in structure formation is to inhibit streamline crossing.
The sound speed
is effectively zero and gas elements can only communicate with one another
via the bulk viscosity with the signal speed.

Bae \& Zhu (2018a) demonstrated that in the linear limit (Goldreich
\& Tremaine; Ogilvie \& Lubow) and in the absence of self-gravity and dissipative processes
the phase of spiral arms agree with hydrodynamic simulations; moreover, they point out
kinematic viscosity is not {explicitly} included in the simulations.  In their case
disturbances communicate with the explicit sound speed, whereas in the present simulations,
the sound speed
is effectively zero.  The results here would appear to be relevant to a cold disk
with vanishing disk thickness.  In other words, the mass of the
planet is much higher than the thermal mass (that at which the Hill or
accretion radius is equal to the vertical scale height), even if we take the sound speed
to be the signal speed with bulk viscosity. The implication is that within this limit
the discussion in terms of periodic orbits should be valid.

To demonstrate the effect of bulk viscous dissipation I have run a set of simulations where
the dissipation timescale, $t_s$ (Eqs. 1 and 2) is varied.  This parameter determines
the strength of the viscous interaction with larger values implying
a smaller effect of dissipation.  Figure 7 shows the results of these
simulations;   panels $a$ to $d$ correspond respectively to
a dissipation timescale of 0.2, 1, 5. 25 time units (one unit is scaled
to 1 AU/1 km/s or 4.75 years for the 4 $M_J$ case).  For the strongest
dissipation (case $a$) the spiral structure is present, but the arms
are rather diffuse and broad.  Case $b$ ($t_s=1$), the level of dissipation
is that
considered in the previous simulations (Fig.\ 3) and in case $c$ the dissipation is much
weaker ($t_s=5$).  Here we
see the arm connecting to the planet is less conspicuous, but a third
arm with lower pitch angle is more prominent.  Finally, in case $d$ ($t_s=25$)
the spiral structure consists of two tightly wound arms just inside
the inner resonance and begins to resemble a complete ring.  In
the limit of vanishing dissipation the structure is that of ring-gap rather
than spirals.  These simulations illustrate, at least in this limit of
cold thin disks, that the kinetic bulk viscosity is necessary for the
development of a spiral structure excited by a massive planet, and that the properties
of the arms depend upon the magnitude of the viscosity.

\section{Extended disks and non-circular planetary orbits}

In all of the examples considered above, the radius of the circular planetary
orbit was near the outer edge of the gaseous disk ($r_p = 32$, $R_d = 40$).
Here I consider the response of the gas when the
planet is more deeply embedded within the disk.  Such a case is shown
in Fig. 7 where $R_p=28$ and $R_d = 50$, and   where $M_p = 4$;
the disk extends beyond
the outer-Lindblad resonance at a  radius of 37.

\begin{figure}
\begin{subfigure}[b]{0.6\textwidth}
\includegraphics[height=5cm]{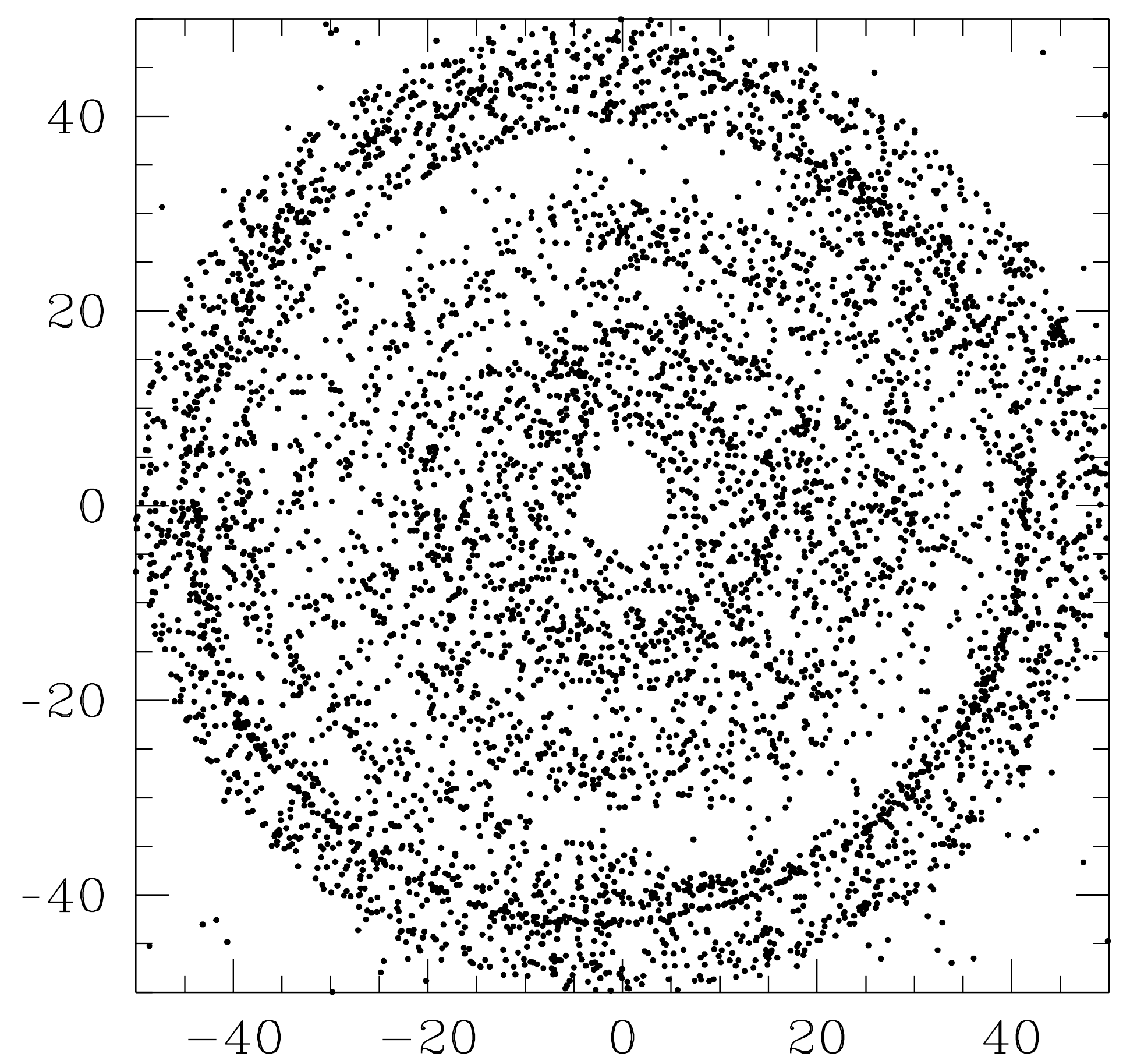}
\captionsetup{margin=10pt,font=small,labelfont=bf}
\caption{Dissipationless particle distribution}
\end{subfigure}
\begin{subfigure}[b]{0.6\textwidth}
\includegraphics[height=5cm]{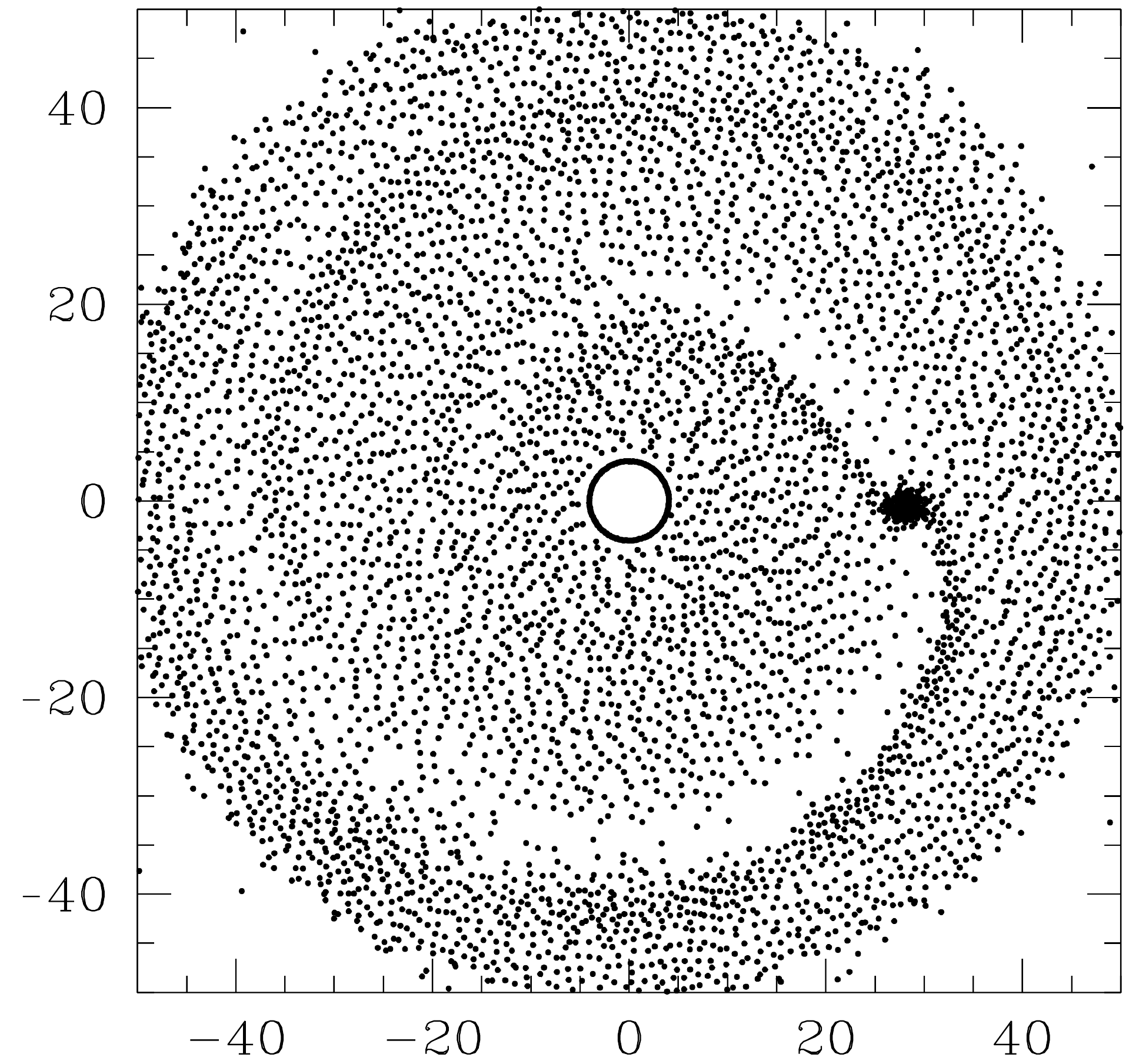}
\captionsetup{margin=10pt,font=small,labelfont=bf}
\caption{Corresponding gas density distribution.}
\end{subfigure}
\caption{Particle and gas distribution for disks extending
beyond the outer resonance (r=42).}
\end{figure}

\vskip 1\baselineskip
In the first panel (Fig. 7a) we see the response of the ensemble of 6000
particles with no dissipation (pure dynamical orbits) after
five planetary rotation periods.  This
reflects the basic structure of the periodic orbits; in particular
we see a set of rings and gaps (particularly obvious at the location
of the OLR) that is characteristic of the ballistic simulations.
Figure 7b illustrates the response when dissipation is
added via eq.\ 1. Within the orbit of the planet the
spiral pattern is that seen in the previous simulation (Fig.\ 3c),
but beyond corotation a second
almost disconnected  two-arm spiral is evident;  the principal arm
is the extension of the planetary wake beyond  corotation, and the
secondary arm appears completely disconnected from the inner structure.
This could be viewed as a prediction.  For a disk extending well
beyond the circular planetary orbit of a major planet,   two separate spiral patterns should be observed.
\vskip 1\baselineskip

The properties of the double-spiral system are more obvious on
a polar density map, as shown in Fig.\ 8.
It is particularly clear in this plot that the run of pitch
angles for arms beyond the planet orbit radius differs
from that within corotation in the sense that the pitch angles become
smaller with increasing radius.
Moving away from corotation in either direction, the arms wrap up.
This could be useful in identifying the location of the planet
with respect to the spirals (see also Bae \& Zhu 2018b).

\vskip 1\baselineskip
A second issue concerns deviations from a pure
circular orbit.  What is the form of the spiral pattern excited by
a planet on a more general elliptical orbit?  Figure 9 illustrates
the effect of such deviations for the 4 $M_J$ example.  In both  cases
the  elliptical orbit is broken down into the circular motion of the
guiding center and the epicycle.  In panel $9a$ the radius of the
guiding center is 35 AU and the eccentricity is 0.1.  In panel $9b$
the guiding center is at 38.7 AU and the eccentricity is 0.21.
\begin{figure}
\includegraphics[height=6.cm]{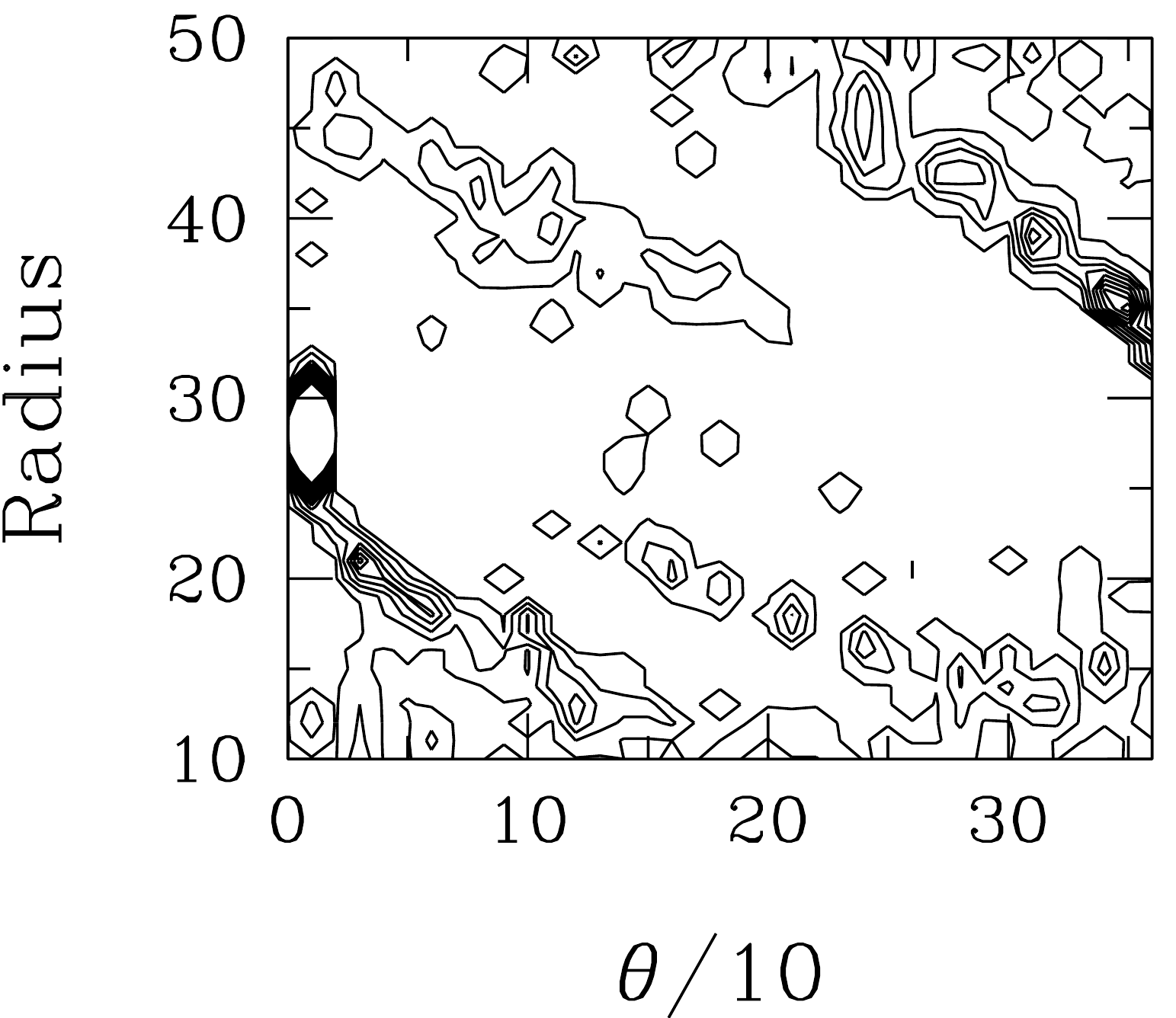}
\captionsetup{margin=10pt,font=small,labelfont=bf}
\caption{Gas surface density distributions in polar projection for disk
extending beyond the outer resonance (Fig. 7a).}
\end{figure}

The basic spiral pattern is essentially identical to that
excited by the planet on the circular orbit at the guiding
center, but the structure of the primary arm that connects to the planet
differs and is transient. This is also generally true for more eccentric orbits
(eccentricity of 0.4 for example), but additional transient structures appear
at corotation.  I have not pursued this further because
of the certainty of secular evolution of the planetary orbit due
to gravitational interactions with the excited structure (orbital
decay),  but the conclusion is that the basic
spiral pattern generated is almost independent of deviations from
circular motion, expected because the basic pattern
of underlying periodic orbits depends on the radius of the
guiding center.

\vskip 1\baselineskip
\begin{figure}
\begin{subfigure}[b]{0.6\textwidth}
\includegraphics[height=5cm]{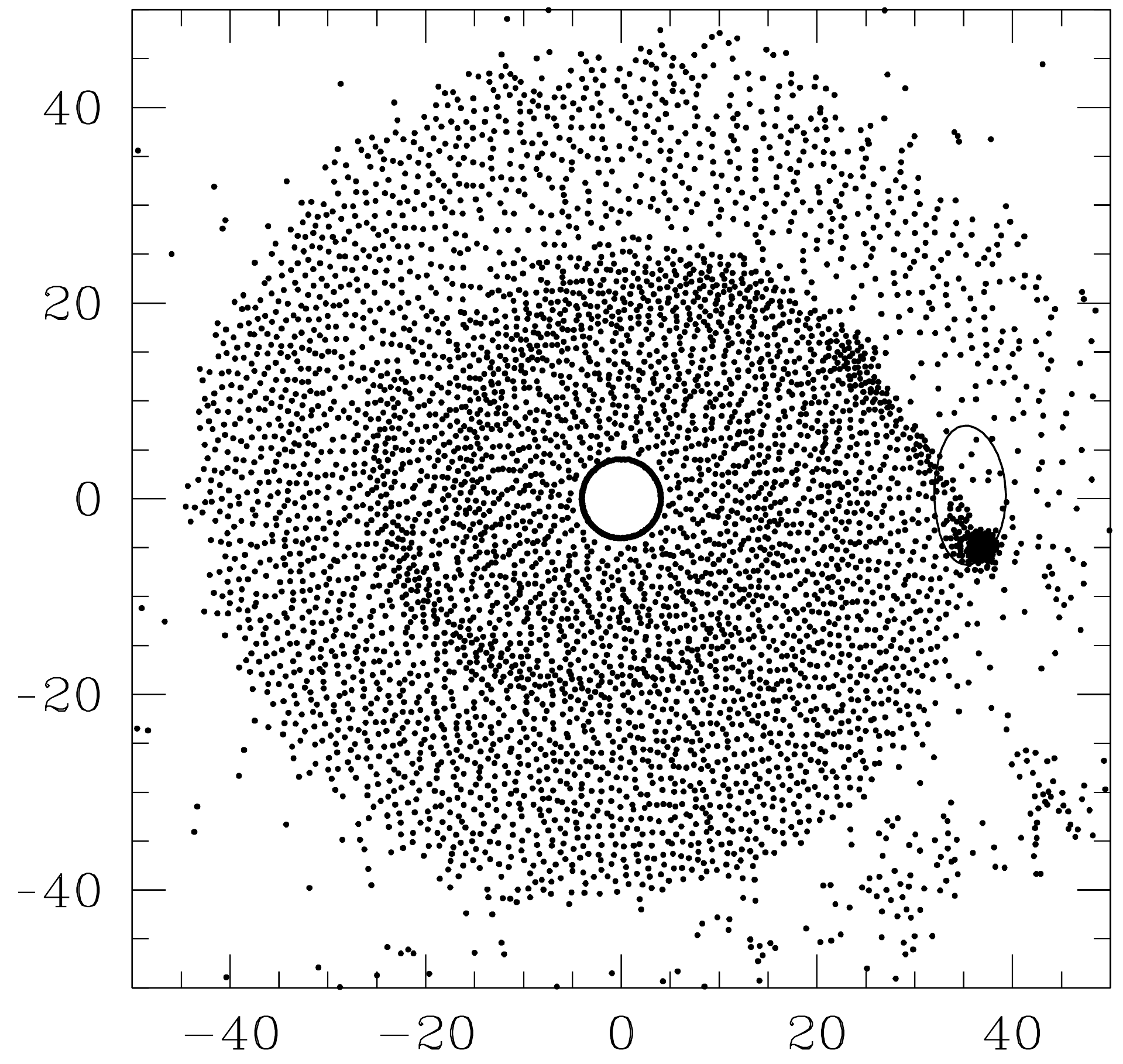}
\captionsetup{margin=10pt,font=small,labelfont=bf}
\caption{Eccentricity = 0.1.}
\end{subfigure}
\begin{subfigure}[b]{0.6\textwidth}
\includegraphics[height=5cm]{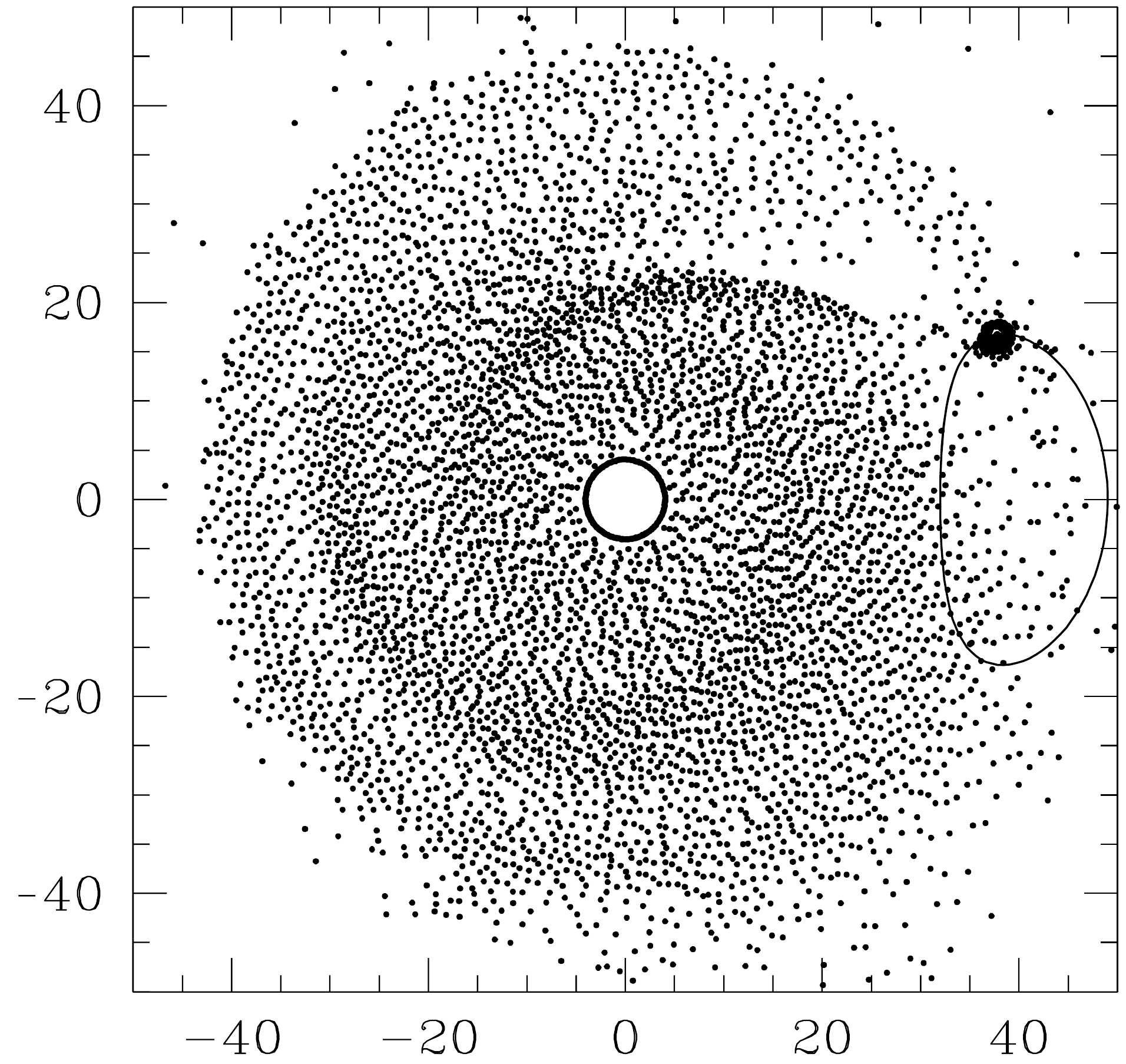}
\captionsetup{margin=10pt,font=small,labelfont=bf}
\caption{Eccentricity = 0.21.}
\end{subfigure}
\caption{Gas density distribution for 4 $M_J$ planets on non-circular
orbits.  The epicycle about the guiding center is shown in both cases.}
\end{figure}

\section{Direct comparison with observations}

As emphasized above, with respect to the observable characteristics of the generated
spiral structure the results here are generally consistent with
the sophisticated simulations that include more physical
effects.  Therefore, one could say that insofar as  the previous calculations
(such as those of Zhu et al. 2015) agree with the observations, so
do these simplified simulations.  Nonetheless, detailed comparisons
of the results here with the observed spiral structure in actual
PPDs, particularly with respect to morphology, remain of interest.
How well do the shapes of the spirals formed by overlapping of
periodic orbits in the presence of dissipation agree with observations?

Here I consider one specific case, SAO 206462 (HD 135344B), because it presents
a reasonably clean two-arm spiral with low inclination to the line of sight
(on the order of 11 degrees), and it is clearly a case where
the mass of the PPD is much less than that of the central star;
self-gravity of the disk probably plays no role in exciting or maintaining the
observed structure (van der Marel et al. 2016).
Moreover, the spiral structure has the general characteristics (Fig.\ 6)
typical
of objects of this kind.  The spirals sweep through
about 200 degrees covering a radial range of 30 to 80 au and
the mass of the central star is estimated to be 1.7 M$_\odot$. To produce
a spiral of this scale requires a planetary orbit having a semi-major
axis  of roughly
80 au and a planetary mass in excess of 0.002 that of the star
(.0034 M$_\odot$).  I have simulated several such cases with
these properties for three rotation periods over a range of
orbital eccentricities.  One of these is shown in Fig.\ 10 along with
the observed system  in the near-infrared J band (Stolker et al. 2018).
\onecolumn
\begin{figure}
\includegraphics[height=12cm]{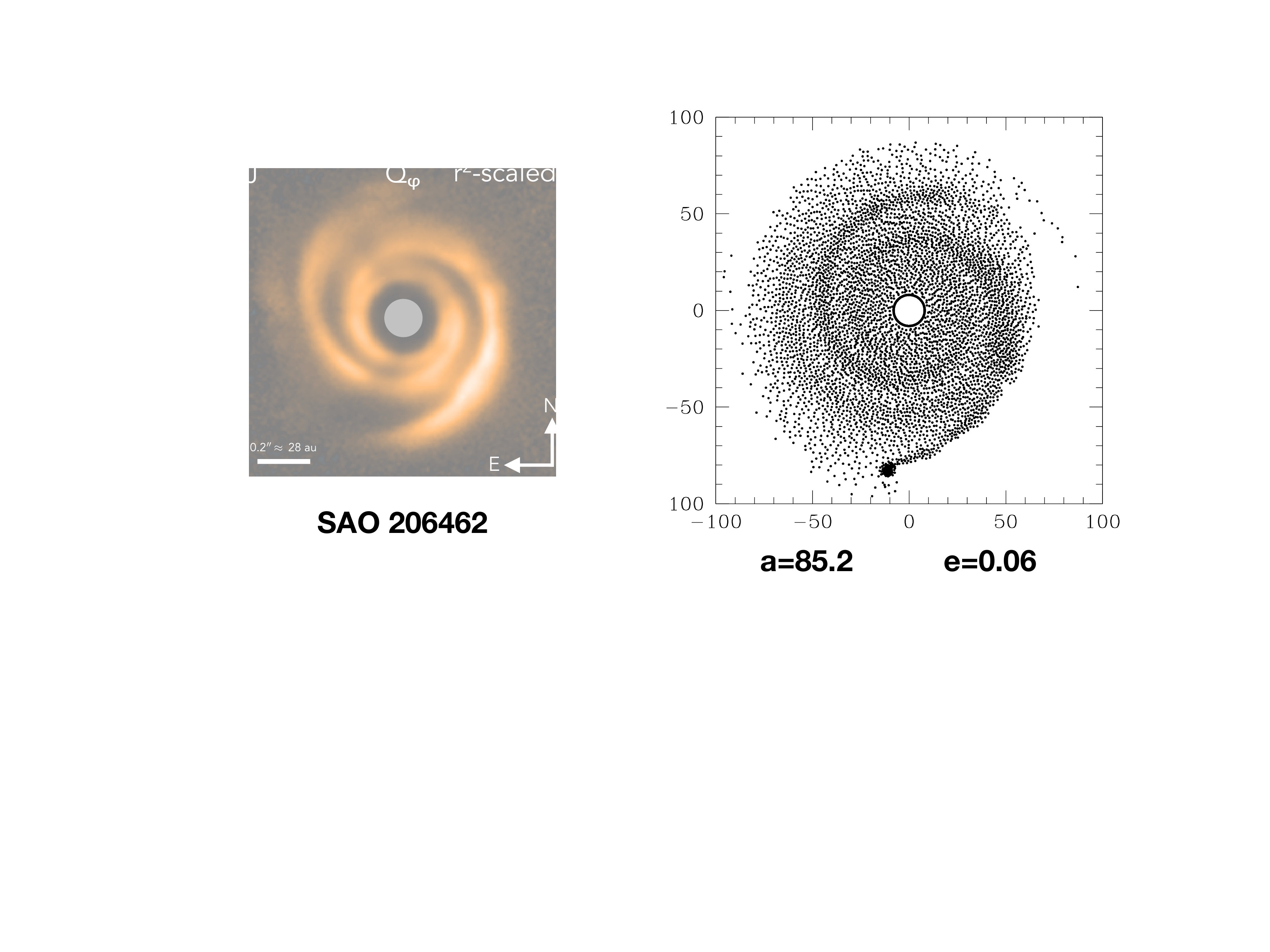}
\captionsetup{margin=10pt,font=small,labelfont=bf}
\caption{ Left panel: Structure observed in SAO 206462
in the J-band polarized continuum with $r^2$ scaling (with permission of
Stolker et al. 2016).
The small bar on the
lower left shows a linear scale of 28 AU at the assumed distance to
the object of 140 pc.  Right panel: Snapshots of
the distribution
of 6000 sticky particles in orbits about a 1.7 $M_\odot$
star perturbed by a 0.0068 $M_\odot$ planet on orbit with semi-major
axis of nearly 85 AU and an eccentricity of 0.061.}
\end{figure}

The simulation is for a planet of 0.004 $M_*$ (0.0068 $M_\odot$)
on an elliptical orbit (eccentricity of 0.061)
at a distance of 85 au to the star. Figure\ 10
is a snapshot after approximately three planetary rotations.
The simulation provides
a reasonable representation of the observed structure. In particular
it demonstrates that the spiral arms wrap into a ring at about
0.4 times the planet orbital semi-major axis and that the bisymmetry is
broken by the arm connecting to the spiral.  In general
this two-dimensional simulation with orbital motion in the
presence of a simple scheme for viscosity adequately describes
the phenomenon.

\section{Conclusions}

In an axisymmetric potential perturbed by a weak rotating bisymmetric
distortion two families of periodic orbits elongated perpendicular to
one another appear near the inner-Lindblad resonance.
These orbits intersect and therefore cannot represent gas stream lines;  
viscosity forces gas flow onto gradually rotating stream lines, and the 
resulting density distribution
is  spiral in form,  hence the appearance of spiral structure in a
pure bar-like potential.

The same effect occurs in a disk rotating in a Keplerian
potential perturbed by a planet even though the perturbation is not bisymmetric.
This occurs because the same two orbit families arise near the inner resonance (2-1 resonance)
and the overlap of these families becomes significant when the mass of the satellite
is more than 1/1000 of the central mass.  Hence, the appearance of a spiral structure is
inevitable in a sufficiently dissipative gas and dust disk about a solar mass star
perturbed by a planet of one Jupiter mass or higher.  

In the calculations described here the only physical effects included are
two-dimensional orbital motion in a perturbed Keplerian potential combined
with a simple mechanism for an artificial bulk viscosity.  This is all that is necessary for
the generation of spiral structure in the thin disk environment of the forming planetary
system.  There is no radiative transfer, no third dimension, no two-fluid
hydrodynamics, no shadowing, and no self-gravity of the disk.  That is not to 
say that such effects are absent or play no role, but that the form of
the periodic orbits near resonance provides the essential ingredient
for the spiral response of a dissipative medium in a perturbed point-mass field.
This sort of insight can be lost in more complicated calculations that do 
include these additional effects.
The principal goal of the
present work has been to gain some understanding of the underlying mechanism
of spiral structure generation in planet-disk interactions,  to elucidate
periodic orbits as the foundation of such a structure.  In that sense,
this work is essentially a ``proof of concept''.

I cannot claim that all spiral structure observed in PPDs is solely due to
this mechanism of streamline crowding at resonances.  The maximum pitch angle
is limited to roughly 25 degrees which may be exceeded
in some observed systems.  The total range of wrapping is restricted to  about 200 degrees,
and there will inevitably be a breaking of bisymmetry in the observed spiral
structure particularly near the perturbing planet.  In some systems,
particularly those with disk masses approaching that of the central star, a non-Keplerian
potential as well as self-gravity
in the disk will certainly play a role in exciting and shaping the observed structure,
with or without the presence of a massive planet, but it seems to be the case that
in many PPDs the disk mass is a small fraction of the stellar mass and
that the disks are likely to stable against the self-gravitational formation of structure
(Kama et al. 2016).  Even if these conditions are met, it may require a cold,
very thin disk to generate the observed structure via this mechanism.

Massive planets do form in
such disks and they  have the effect of exciting spiral structure.  When the
extent of the disk exceeds the outer-Lindblad resonance, there may even be 
the appearance of a second set of apparently disconnected spiral arms as
in Fig. 8. The inner spiral should wrap into
a circular ring at about 40\% of  the planetary orbit radius; that is to say, the
planet would be found at a distance of roughly 2.5 the ring radius near the maximum
extension of the strongest arm with the largest pitch angle.
The acid test for this mechanism would be
the actual observation of the perturbing (proto)planet at this location.

\begin{acknowledgements}  I am grateful to Renzo Sancisi for very useful comments on
an earlier version of this manuscript and to Tomas Stoker for permission
to use the near-infrared map of SAO 206462 as well as a useful comment
on the broken 180 degree symmetry of the observed spiral structure.  I also
acknowledge numerous very helpful remarks from an anonymous referee, in particular
suggesting evaluating of the
significance of viscosity by simulations with various dissipation timescales
(Fig. 7).
\end{acknowledgements}

\end{document}